%% file: main.tex
\documentclass[%
 aps,
 prb,
 amsmath,amssymb,
 reprint,
 superscriptaddress,
 longbibliography,
 floatfix,
]{revtex4-2}
\usepackage{xcolor}
\usepackage{graphicx}
\usepackage{dcolumn}
\usepackage{bm}

\usepackage{float}
\usepackage{mathtools}

\usepackage[utf8]{inputenc}
\usepackage[T1]{fontenc}
\usepackage{newtxtext}
\usepackage{newtxmath}
\usepackage{physics}
\usepackage{listings}
\usepackage{tikz}
\usetikzlibrary{arrows.meta,positioning,calc}
\usepackage[subpreambles=false]{standalone}
\makeatletter
\@ifpackageloaded{hyperref}{
  \hypersetup{
    colorlinks=true,
    linkcolor=blue,
    citecolor=blue,
    urlcolor=blue
  }
}{
  \usepackage[
    colorlinks=true,
    linkcolor=blue,
    citecolor=blue,
    urlcolor=blue
  ]{hyperref}
}
\makeatother
\setcitestyle{numbers,square,sort&compress}

\usepackage[normalem]{ulem}

\input{newcommands}

\definecolor{darkgreen}{RGB}{0, 100, 0}

\definecolor{dark}{RGB}{100, 0, 0}

\begin{document}

\title[]{Bethe-Salpeter Equation Calculations Based on Numerical Atomic Orbitals and Norm-Conserving Pseudopotentials: Dual-${\bm k}$-Mesh Strategy}
\author{Ziqing Guan}
\affiliation{Institute of Physics, Chinese Academy of Sciences, Beijing 100190, China}

\affiliation{University of Chinese Academy of Sciences, Beijing 100049, China}

\author{Yu Cao}
\affiliation{Institute of Physics, Chinese Academy of Sciences, Beijing 100190, China}
\affiliation{HEDPS, CAPT, School of Physics and College of Engineering, Peking University, Beijing 100871, China}

\author{Min-Ye Zhang}
\affiliation{Institute of Physics, Chinese Academy of Sciences, Beijing 100190, China}
\affiliation{The NOMAD Laboratory at the Fritz-Haber-Institut der Max-Planck-Gesellschaft, Berlin 14195, Germany}

\author{Peize Lin}
\affiliation{Institute of Artificial Intelligence, Hefei Comprehensive National Science Center, Hefei 230026, Anhui, China}

\author{Ruiyi Zhou}
\affiliation{Institute of Molecular Physical Science, ETH Zürich, 8093 Zürich, Switzerland}
\affiliation{Department of Chemistry, University of North Carolina at Chapel Hill, Chapel Hill, North Carolina 27599, USA}

\author{Xinguo Ren}
\email{renxg@iphy.ac.cn}
\affiliation{Institute of Physics, Chinese Academy of Sciences, Beijing 100190, China}

\date{\today}

\begin{abstract}
	We present a dual-$\bm k$-mesh implementation of the Bethe--Salpeter equation (BSE) based on numerical atomic orbitals (NAOs) and norm-conserving pseudopotentials within the ABACUS+LibRPA framework. By exploiting the localized resolution-of-identity (LRI) technique, the screened Coulomb interaction is cast into a real-space, unit-cell-indexed form $W_{\mu\nu}(\bm R)$ on the Born--von~K\'arm\'an supercell of the coarse $\bm k$-mesh. This real-space representation enables a Fourier interpolation of the BSE kernel from the coarse $\bm k$-mesh used in the preceding $GW$ calculation to an arbitrarily dense $\bm k$-mesh on which the BSE Hamiltonian is assembled and diagonalized, thereby giving rise naturally to a dual-$\bm k$-mesh workflow. Building on this scheme, we systematically examine the convergence of the absorption spectra with respect to the NAO basis set, the auxiliary basis set, and the $\bm k$-point sampling. Benchmark calculations for both molecular and periodic systems collectively validate the accuracy of the present implementation and establish the dual-$\bm k$-mesh strategy as a practical approach for $GW$+BSE calculations.
\end{abstract}

\maketitle

\section{Introduction}
The quantitative prediction of optical excitations in condensed-matter and molecular systems remains a central task in first-principles electronic-structure theory. In semiconductors, insulators, and low-dimensional materials, the optical response is often governed by excitonic effects arising from the correlated motion of photoexcited electrons and holes. Although time-dependent density functional theory can be highly effective for certain classes of systems, its accuracy for charge-transfer excitations and strongly bound excitons is often limited by the approximate exchange--correlation kernel \cite{Onida2002,Maitra_2017}.  In this context, many-body perturbation theory based on the one-particle Green's function provides a systematic framework, in which quasiparticle energies are described at the $GW$ level and neutral excitations are obtained from the Bethe--Salpeter equation (BSE) \cite{Salpeter1951,Strinati1988,Onida2002}. The resulting $GW$+BSE scheme is now widely regarded as one of the most reliable first-principles approaches for optical spectra and excitonic properties.

Over the past decade, the BSE methodology has advanced substantially along several fronts. Its range of application has expanded from bulk solids~\cite{Rohlfing2000} to molecular systems~\cite{Jacquemin2015,Blase2020} and heterogeneous interfaces~\cite{Okada2018}. At the same time, considerable effort has been devoted to reducing the computational cost of solving the BSE Hamiltonian, including non-uniform Brillouin-zone (BZ) sampling strategies~\cite{Alvertis2023}, modern large-scale algorithms based on GPU acceleration~\cite{Yu2024}, stochastic sampling on the screened electron–hole interaction kernel~\cite{Bradbury2023}, and energy-specific BSE implementation by restricting the Davidson subspace expansion to predefined energy windows~\cite{Hillenbrand2025}.  Alongside these computational developments, theoretical formulation has also made huge progress. Attaccalite \textit{et al.}~\cite{Attaccalite2011} developed a real-time-propagation BSE formalism that extends the method beyond the linear-response regime, and many researchers subsequently developed similar time propagation approaches and applied them to monolayer transition metal dichalcogenides~\cite{JiangSA2021,Chan2021}, moiré heterobilayers~\cite{Hu2023,ChangLee2024}, and molecular systems~\cite{Marek2025}. Complementarily, Ruan \textit{et al.}~\cite{Ruan2024} developed a frequency-domain nonlinear optical response for BSE calculations. To reduce the dependence on the underlying DFT starting point, Li \textit{et al.}~\cite{Li2022} proposed a renormalized-singles $GW$ approach for BSE, which has demonstrated excellent accuracy for molecular excitations. Beyond valence optical excitations, which are the focus of the present work, BSE methods have long been applied to X-ray absorption and inelastic X-ray scattering~\cite{Shirley1998,Vinson2011}. Recent studies have further extended these core-level applications to complex systems~\cite{Tang2022,Urquiza2024,Nicolaou2025}. Alongside this expanding scope, space-group symmetry offers another route to reducing the computational cost of BSE calculations. Bajaj \textit{et al.}~\cite{bajaj2025}, Nalabothula~\textit{et al.}~\cite{Nalabothula2026} and Stöhler~\textit{et al.}~\cite{stöhler2026} have exploited space-group symmetries in the construction and block-diagonalization of the BSE matrix and in the subsequent characterization of excitonic states. Together, these developments show that contemporary BSE research is driven not only by predictive accuracy, but also by scalability, algorithmic efficiency, and transferability across different classes of materials.

Despite this progress, the convergence of BSE spectra with respect to $\bm k$-point sampling remains a major bottleneck for periodic systems. The difficulty originates from the complexity of the wavefunction in reciprocal space, which necessitates dense Brillouin-zone sampling to construct the BSE kernel and achieve converged excitation energies and oscillator strengths. Conventional uniform Monkhorst--Pack $\bm k$-meshes therefore incur a rapidly increasing computational cost, making systematic convergence studies expensive and, in some cases, impractical. To address this bottleneck, dual-$\bm k$-mesh schemes that interpolate from a coarse to a dense $\bm k$-mesh have been proposed: Rohlfing and Louie~\cite{Rohlfing2000} interpolated the BSE matrix elements by projecting wavefunctions from the coarse onto the dense grid; Kammerlander~\cite{Kammerlander2012} applied Wannier interpolation to the non-interacting two-particle correlation function while retaining the coarse $\bm k$-mesh for the interaction kernel; Gillet~\cite{Gillet2016} interpolated the interaction kernel on the dense grid using the eight nearest $\bm k$-points from the coarse grid; and Alliati~\cite{Alliati2022} introduced the so-called ``diagonal kernel extension'' method, in which the fine $\bm k$-points are grouped into domains centered at the coarse $\bm k$-points, and each coarse-grid kernel matrix element is extended to a block in the fine-grid matrix by copying its value to the diagonal entries of that block while setting all off-diagonal entries to zero.

In this work, we also develop a dual-$\bm k$-mesh strategy, where the interaction kernel (especially screened Coulomb interaction) is represented in a real-space form, which enables Fourier interpolation from the coarse $\bm k$-mesh used for $GW$ to a dense $\bm k$-mesh for BSE, naturally giving rise to a dual-$\bm k$-mesh workflow. Our implementation builds on the work of Zhou \textit{et al.}~\cite{Zhou2025}, who established the BSE formalism within the framework of numerical atomic orbital (NAO) basis sets and the localized resolution-of-identity (LRI) technique, thereby providing the formal foundation for Fourier interpolation in the localized real-space representation. While the original implementation resides in the all-electron DFT package FHI-aims\cite{blum2009ab,blum2026roadmap}, this work is developed within the pseudopotential-based ABACUS+LibRPA framework. ABACUS~\cite{ABACUS2016,ABACUS2024,ABACUS2025} provides DFT eigenstates as a starting point, and LibRPA serves as a post-DFT framework for many-body Green's function calculations in the NAO representation. Within LibRPA, the random-phase approximation (RPA) correlation energies~\cite{Shi2024,Shi2025} and $GW$ quasiparticle energies~\cite{Zhang2026, jia2026, gong2026} have been implemented previously, establishing a solid foundation for the present work. Our BSE implementation has three main features: First, it is formulated in the NAO framework and naturally exploits orbital locality. Second, the LRI technique provides an efficient route for treating the screened interaction in a compact auxiliary basis. Third, the real-space formulation enables a clean separation between the coarser $\bm k$-mesh used to construct the screened interaction and the denser $\bm k$-mesh used in the excitonic Hamiltonian, which facilitates dual-$\bm k$-mesh interpolation and practical convergence acceleration while maintaining close agreement with established BSE implementations. With these features, we assess the accuracy of our BSE implementation through systematic convergence tests and benchmark calculations, including absorption spectra and exciton binding energies for a range of prototypical materials.

The remainder of this paper is organized as follows. In Sec.~\ref{sec:Theory}, we outline the theoretical framework, beginning with Green's function theory and deriving the BSE formalism. Sec.~\ref{sec:Details} discusses the implementation details in ABACUS+LibRPA, including the LRI technique in the NAO framework and the dual-$\bm k$-mesh strategy for Fourier interpolation. Sec.~\ref{sec:Results} presents a systematic analysis of the convergence behavior with respect to the NAO basis set, the auxiliary basis set, and the $\bm k$-point sampling, together with benchmark results against standard BSE implementations and available experimental data. 
Finally, we summarize our findings and provide an outlook in Sec.~\ref{sec:Conclusion}.

\section{Theory} \label{sec:Theory}

\subsection{Green's function theory}

The theoretical foundation of our approach rests on the single-particle Green's function, which is formally defined as
\begin{equation}
	G(1,2) = -i\langle N,0|T[\hat \psi(1) \hat \psi^\dagger(2)] |N,0\rangle\, ,
\end{equation}
where $1\equiv(\bm r_1,t_1)$ represents a combined space-time coordinate and $|N,0\rangle$ denotes the $N$-electron ground state. In practical implementations, the non-interacting Green's function $G^0$ is obtained with Kohn-Sham orbitals and orbital energies. Within the NAO basis set formalism, the Kohn-Sham eigenstates are expressed as a linear combination of localized basis functions
\begin{equation}
	\psi^{\bm k}_{n}(\bm r) = \sum_{s, \bm R} c_{sn}^{\bm k} \phi_s(\bm r- \bm R-\bm \tau_s) e^{i \bm k \cdot \bm R}\, ,
\end{equation}
where $\phi_s$ represents a localized atomic orbital centered at $\bm R +\bm \tau_s$  ($\bm R$ being a lattice vector and $\bm \tau_s$ the position vector in the unit cell) and $c_{sn}^{\bm k}$ denotes the corresponding expansion coefficient.

To leverage the intrinsic locality of the NAO basis, we adopt a real-space imaginary-time representation of the Green's function
\begin{equation}
	\begin{aligned}
	G^0(\bm r, \bm r',i\tau) = & \sum_{s,t} \sum_{\bm R,\bm R'} G^0_{st} (\bm R'-\bm R, i\tau)\\
	& \phi_s(\bm r-\bm R -\bm \tau_s)\phi_t(\bm r'-\bm R' -\bm \tau_t)\, ,
	\end{aligned}
\end{equation}
where the matrix elements in the localized basis assume the following form
\begin{equation}
\begin{aligned}
	&G^0_{st} (\bm R,i\tau) \\
    =& \begin{cases}
		\displaystyle-\frac{i}{N_{\bm{k}}}\sum_{n,\bm{k}}f_{n\bm{k}}c_{sn}^{\bm{k}}c_{tn}^{\bm{k}*}e^{-i\bm{k}\cdot\bm{R}}e^{-(E_{n}^{\bm{k}}-\mu)\tau},\tau\leq0    \\
		\displaystyle\frac{i}{N_{\bm{k}}}\sum_{n\bm{k}}(1-f_{n\bm{k}})c_{sn}^{\bm{k}}c_{tn}^{\bm{k}*}e^{-i\bm{k}\cdot\bm{R}}e^{-(E_{n}^{\bm{k}}-\mu)\tau},\tau>0 &
	\end{cases} \, .
\end{aligned}
\end{equation}
Here, $f_{n\bm{k}}$ denotes the occupation number, $E_n^{\bm{k}}$ represents the Kohn-Sham eigenvalue, and $\mu$ is the chemical potential. Furthermore, $N_{\bm{k}}$ is the number of $\bm k$-points in the first Brillouin zone. This formulation explicitly exploits the spatial locality inherent in the NAO basis, rendering it particularly amenable to real-space treatments of many-body quantities. The locality-based strategy has been successfully implemented within the LibRPA framework for the computation of RPA correlation energies~\cite{Shi2024,Shi2025} and $GW$ quasiparticle quantities~\cite{Zhang2026,jia2026,gong2026}.

Based on $G^0$, the non-interacting response function $\chi^0$, the dielectric function $\varepsilon$, the screened Coulomb interaction $W$, and the exchange-correlation (XC) part of the self-energy $\Sigma_\text{xc}$ are calculated sequentially as follows:
\begin{equation}
	\begin{aligned}
		\chi^0(1,2) & = -i G^0(1,2) G^0(2,1)\\
		\varepsilon(1,2) & = \delta(1,2)- \int \mathrm{d}3 v(1,3) \chi^0(3,2) \\
		W(1,2) & = \int \mathrm{d}3 \varepsilon^{-1}(1,3) v(3,2) \\
		\Sigma_\text{xc}(1,2) & = i G^0(1,2) W(2,1)\, ,
	\end{aligned}
\end{equation}
Determining the self-energy $\Sigma_\text{xc}$ via the above equations is the so-called $G^0W^0$ approach, which provides a starting point for the subsequent calculations based on the BSE, to be discussed below.

Our $G^0W^0$ implementation preceding BSE is carried out within the NAO basis set framework, combined with the space-time formulation and the LRI technique. This implementation features a $O(N^2)$ scaling behavior for the calculations of the density response function $\chi^0$ and self-energy $\Sigma_\text{xc}$, thus enabling the treatment of large, periodic systems. Further details of our $G^0W^0$ implementation can be found in Refs.~\cite{Zhang2026,gong2026}.

\begin{figure*}[htbp]
	\includegraphics[width=0.8\linewidth]{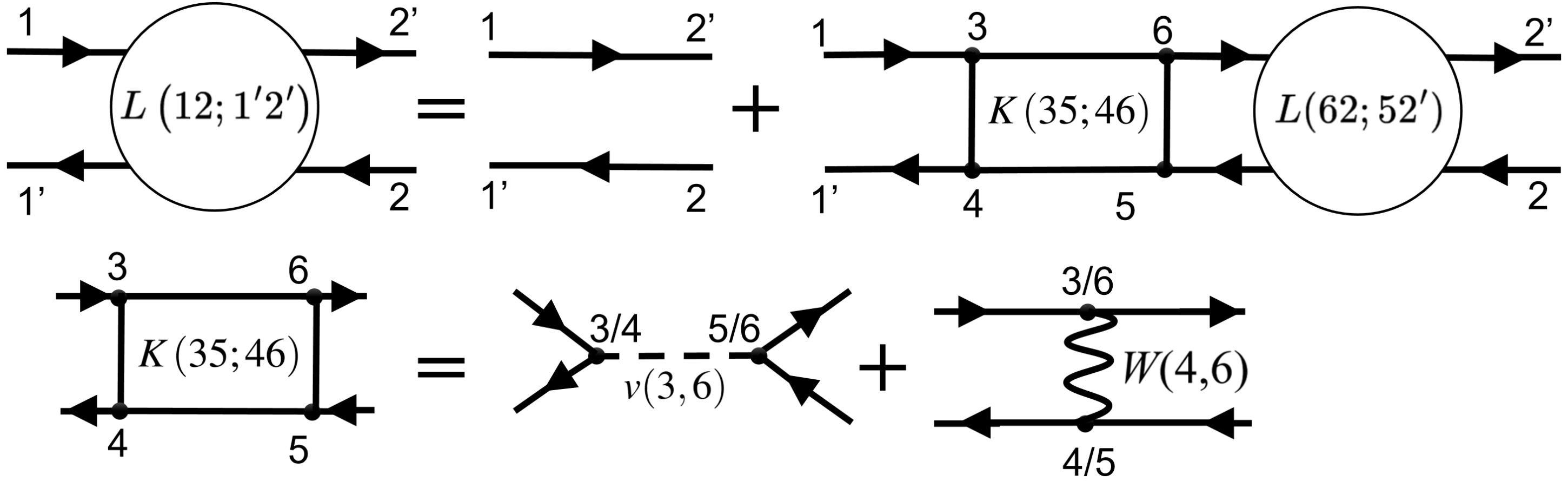}
	\caption{The Feynman diagram of BSE. Upper panel: The bubbles correspond to the full correlation function $L$. The first term on the right-hand side consists of two fermion lines, representing the noninteracting electron-hole correlation function $L^0$. Lower panel: The ladder part labeled by the kernel $K$ is composed of two terms--the bare Coulomb exchange term $v$ and the screened Coulomb direct term $W$.}
	\label{fig:FeynmanBSE}
\end{figure*}

\subsection{Bethe-Salpeter equation}

Following Strinati~\cite{Strinati1984,Strinati1988}, the BSE can be written as a Dyson equation
\begin{equation}\label{Dyson}
	\begin{aligned}
		L\left(12;1^{\prime}2^{\prime}\right)=& L^{0}\left(12;1^{\prime}2^{\prime}\right)+\int\mathrm{d}\left(3456\right)L^{0}\left(14;1^{\prime}3\right) \\
		& \times K\left(35;46\right)L(62;52^{\prime}),
	\end{aligned}
\end{equation}
where $L$ denotes the full correlation function, $L^0$ is the non-interacting two-particle correlation function defined as
\begin{equation}
	L^0(12;1'2') = G^{0}(1,2') G^{0}(2,1')\, ,
\end{equation}
and $K(35;46) = \delta \Sigma_\text{Hxc}(3,4) / \delta{G(6,5)}$ represents the interaction between the electron-hole pairs. It contains the functional derivatives of the Hartree contribution $\Sigma_\text{H}(3,4) = -i \delta(3,4) \int \mathrm d 7 v(3,7)G(7,7^+)$ and of the exchange-correlation self-energy $\Sigma_\text{xc}$. In the $GW$ approximation, neglecting the vertex term, the latter is $\Sigma_\text{xc}(3,4) = i G(3,4) W(4,3)$, and one thus obtains
\begin{equation}\label{kernel}
	K(35,46) = \underbrace{- i v(3,6) \delta(3,4)\delta(5,6)}_\text{bare Coulomb exchange term} + \underbrace{i W(3,4) \delta(3,6) \delta(4,5)}_\text{screened Coulomb direct term}\, .
\end{equation}

Fig. \ref{fig:FeynmanBSE} provides a diagrammatic representation of the BSE. The first term in Eq.~\eqref{kernel} interchanges the electron and hole lines and is therefore referred to as the ``exchange term''. By contrast, the second term preserves the electron-hole character and is denoted as the ``direct term''. It is important to note that, in such a two-particle scattering process, the ``exchange'' term here derives from the Hartree potential (usually referred to as the direct term in the one-particle process), while the
``direct'' term here derives from the dynamical one-electron exchange-correlation potential given by $GW$.  Throughout this work, we adopt the static approximation, i.e., we neglect the frequency dependence of $W(\omega)$ and evaluate it at $\omega=0$.


\subsection{The matrix form of BSE}
To transform the Dyson equation from its integral form \eqref{Dyson} into an eigenvalue equation, it is convenient to express it in the basis of two-body product wavefunctions. In the frequency domain, the non-interacting electron-hole correlation function is given by
\begin{equation}\label{L0}
\begin{aligned}
	&L^0(\bm r_1 \bm r_2;\bm r_1'\bm r_2';\omega)\\
    =&i \sum_{op}(f_p-f_o)\frac{\psi_o(\bm r_1)\psi^*_p(\bm r_1')\psi^*_o(\bm r_2')\psi_p(\bm r_2)}{\omega-(E_o-E_p)+i\eta \operatorname{sign}(E_o-E_p)}\, ,
\end{aligned}
\end{equation}
where $o$ and $p$ label general single-particle states, $f_o$ and $f_p$ are their occupations, and $\eta$ is a positive infinitesimal that ensures the correct analytic structure. Although the full correlation function formally depends on four time variables, it can be reduced to a form that depends on a single frequency variable, assuming that particle creation and annihilation occur simultaneously and that the system possesses time-translation symmetry. 
Under this assumption, the full correlation function can be expressed as
\begin{equation}
\begin{aligned}\label{eq:L_matrix1}
	L(\bm r_1\bm r_2;\bm r_1'\bm r_2';\omega)=i \sum_{S}\frac{\chi^{R}_S(\bm r_1, \bm r_1')\chi^{L}_S(\bm r_2', \bm r_2)}{\omega-\Omega_S+i\eta \operatorname{sign}(\Omega_S)}\, ,
\end{aligned}
\end{equation}
in which $S$ denotes the excited states of the system, encapsulating the information about the electron-hole interactions. The upper right labels $R$ and $L$ correspond to the right eigenstates \eqref{eq:right} and left eigenstates \eqref{eq:left}. The poles of the correlation function, $\Omega_S$, are interpreted as the neutral excitation energies of the system. The excitation amplitudes from the ground state $|N,0\rangle$ to the neutral excited state $|N,S\rangle$ are expressed in the two-body wavefunction basis set as 
\begin{equation}\label{eq:amplitude}
	\begin{aligned}
	\chi^R_S(\bm{x},\bm{x'}) &= \langle N,0| \hat\psi(\bm x)\hat\psi^\dagger(\bm x')|N,S\rangle \\
	&=\sum_{op}X^S_{op} \psi_o(\bm x)\psi^*_p(\bm x') + Y^S_{op} \psi_p(\bm x)\psi^*_o(\bm x')\, ,
	\end{aligned}
\end{equation}
where the coefficients of excitation amplitudes $X^S_{op}, Y^S_{op}$ are the expansion coefficients of the two-body product basis set $\{ \psi_o\psi_p^*\}$ and can be obtained by solving an eigenvalue problem \cite{Rohlfing2000, Onida2002}.

Here we briefly derive how to transform BSE into an eigenvalue problem. The non-interacting correlation function \eqref{L0} can be represented as a matrix that contains only diagonal elements (omitting the infinitesimal $\eta$ for simplicity):
\begin{equation}
	L^0_{op,qr} = \frac{f_p-f_o}{\omega-(E_o-E_p)}\delta_{qo}\delta_{rp} \label{L0_c}
\end{equation}
\begin{equation}
\begin{aligned}
	&L^0(\bm r_1,\bm r_2;\bm r_1',\bm r_2')\\
    =&i\sum_{op,qr}L^0_{op,qr} \psi_o(\bm r_1)\psi_p^*(\bm r_1')\psi^*_q(\bm r_2')\psi_r(\bm r_2)\, .
\end{aligned}
\end{equation}

With treatment similar to that of the full correlation function, the Dyson equation \eqref{Dyson} can be transformed into a matrix equation,
\begin{equation}\label{Dyson_c}
	L_{op,qr}=L_{op,qr}^0+\sum_{klmn}L_{op,kl}^0K_{kl,mn}L_{mn,qr}
\end{equation}
where the interaction kernel is represented as
\begin{equation}
	K_{kl,mn}=i\int\mathrm{d}(3456)\psi^*_k(3)\psi_l(4)K(35;46)\psi_m(6)\psi^*_n(5)  \, .
\end{equation}

Eq.~\eqref{Dyson_c} is a matrix Dyson equation in the two-particle transition space. Substituting Eq.~\eqref{L0_c} into Eq.~\eqref{Dyson_c}, we obtain
\begin{widetext}
\begin{equation}
    \sum_{mn}\bigg\{\omega\delta_{om}\delta_{pn}-[\underbrace{(E_o-E_p)\delta_{om}\delta_{pn}+ (f_p-f_o) K_{op,mn}]}_{H^\text{BSE}_{op,mn}}\bigg\} L_{mn,qr}=(f_p-f_o)\delta_{oq}\delta_{pr}\, .
\end{equation}
\end{widetext}
Defining the terms in braces as a two-particle Hamiltonian $H^\text{BSE}$, one can immediately find that
\begin{equation}\label{eq:L_matrix2}
	\begin{aligned}
	L_{op,qr} &= (\omega - H^\text{BSE})_{op,qr}^{-1} (f_r-f_q) \\
	&=\sum_{S} \frac{Z_{op,S} Z_{S,qr}^{-1}}{\omega-\Omega_S} (f_r-f_q)\, ,
	\end{aligned}
\end{equation}
the second line assumes $H^\text{BSE}$ has eigenvalues $\Omega_S$ and right eigenvectors $Z_{op,S}$. $Z_{S,qr}^{-1}$ means the inverse of the right eigenvector matrix, which are also called left eigenvectors. Comparing Eq. \eqref{eq:L_matrix1} and Eq. \eqref{eq:L_matrix2}, one can find the relationship
\begin{equation}
	\begin{aligned}
	\chi^R_S(\bm r_1, \bm r_1') &= \sum_{op} Z_{op,S} \psi_o(\bm r_1)\psi_p^*(\bm r_1')\, , \\
	\chi^L_S(\bm r_2', \bm r_2) &= \sum_{qr} Z_{S,qr}^{-1} (f_r-f_q) \psi_q^*(\bm r_2')\psi_r(\bm r_2)\, .
	\end{aligned}
\end{equation}

The two-body product basis set $\{\psi_o \psi_p^*\}$ is typically chosen as electron-hole pair, and the difference of the occupation numbers appearing in the numerator in Eq.\eqref{L0_c} equals $\pm 1$ when integer occupations are assumed. Following the common convention, we use $i,j$ to denote valence bands (occupied hole states), and $a,b$ to denote conduction bands (empty electron states). Then, in the basis set $\{\psi_a \psi_i^*; \psi_i \psi_a^* \}$ , the BSE Hamiltonian matrix has the following structure:
\begin{equation}
	H^\text{BSE} =
	\begin{bmatrix}A    & B    \\
               -B^* & -A^*
	\end{bmatrix}\, .
\end{equation}
Substituting Eq.~\eqref{kernel} into the BSE Hamiltonian, we obtain
\begin{equation}\label{eq:Hamiltonian}
	\begin{aligned}
		A_{ai \bm k_1,bj \bm k_2}= & (E_{a\bm{k}_1}^{\text{GW}}-E_{i\bm{k}_1}^{\text{GW}})\delta_{ab}\delta_{ij}\delta_{\bm{k}_1\bm{k}_2} \\
		& +\alpha(a\bm{k}_1^*,i\bm{k}_1|V|j\bm{k}_2^*,b\bm{k}_2)\\
		& -(j\bm{k}_2^*,i\bm{k}_1|W|a\bm{k}_1^*,b\bm{k}_2)\\
		B_{ai \bm k_1,bj \bm k_2}= & \alpha(a\bm{k}_1^*,i\bm{k}_1|V|b\bm{k}_2^*,j\bm{k}_2)\\
		& -(b\bm{k}_2^*,i\bm{k}_1|W|a\bm{k}_1^*,j\bm{k}_2)\, ,
	\end{aligned}
\end{equation}
where $\alpha=2$ for the spin singlet channel, $\alpha=0$ for the spin triplet channel. This formula applies only to spin-degenerate systems, as it has been diagonalized in the spin space~\cite{Rohlfing2000}. The parenthetical notations present the integral below
\begin{equation}\label{eq:BSEkernel}
	\begin{aligned}
		& (a\bm{k}_1^*,i\bm{k}_1|V|j\bm{k}_2^*,b\bm{k}_2) \\
		\equiv & \iint \mathrm d\bm r\mathrm d\bm r' \psi_{a}^{\bm k_1*}(\bm r)\psi_{i}^{\bm k_1}(\bm r) V(\bm r, \bm r') \psi_{j}^{\bm k_2*}(\bm r') \psi_{b}^{\bm k_2}(\bm r')\, ,\\
        & (j\bm{k}_2^*, i\bm{k}_1|W|a\bm{k}_1^*, b\bm{k}_2) \\
		\equiv & \iint \mathrm d\bm r\mathrm d\bm r' \psi_{j}^{\bm k_2*}(\bm r)\psi_{i}^{\bm k_1}(\bm r) W(\bm r, \bm r') \psi_{a}^{\bm k_1*}(\bm r') \psi_{b}^{\bm k_2}(\bm r')\, ,
	\end{aligned}
\end{equation}
other notations are defined analogously. It is straightforward to verify that the block $A$ is Hermitian, while the block $B$ is symmetric. 

In this work, both the Tamm-Dancoff Approximation (TDA) and full BSE are implemented. In the TDA, only the resonant block $A$ of the BSE Hamiltonian is retained, while the coupling (anti-resonant) block $B$ is neglected, leading to a Hermitian eigenvalue problem. For the full BSE matrix, we follow the stable paired-eigenvalue formulations and normalization convention as discussed in Refs.~\cite{Shao2016,Penke2020}, which allow direct access to both $X$ and $Y$ amplitudes in a numerically robust manner.
The solutions of this BSE matrix appear in positive-negative pairs $\pm \Omega_S$, whose right eigenvectors are
\begin{equation}\label{eq:right}
	Z^R_+=\begin{bmatrix}X \\
		Y
	\end{bmatrix},\ Z^R_-=\begin{bmatrix}Y^* \\
		X^*
	\end{bmatrix}\, ,
\end{equation}
adopting the normalization convention $X^\dagger X-Y^\dagger Y=\mathbf {I}$, the left eigenvectors are
\begin{equation}\label{eq:left}
	Z^L_+=\begin{bmatrix}X^\dagger & -Y^\dagger	\end{bmatrix},\ Z^L_-=\begin{bmatrix}-Y^T & X^T	\end{bmatrix}\, .
\end{equation}

Within the TDA, the BSE spectrum is calculated through the imaginary part of the macroscopic dielectric function
\begin{subequations}\label{eq:abs}
\begin{align}
	\varepsilon_{2}(\omega) & \overset{\text{vel}}{=}\sum_{S}\frac{4\pi^{2}}{3N_{\bm{k}}V}\left|\sum_{ai\bm{ k }}\frac{\langle i\bm{k}|\hat{\bm v} |a\bm{k}\rangle}{E_{a\bm{k}}-E_{i\bm{k}}} X_{ai\bm{k}}^{S}\right|^{2}\delta(\omega-\Omega_{S})
	\label{eq:abs_vel} \\
	& \overset{\text{len}}{=}\sum_{S}\frac{4\pi^{2}}{3N_{\bm{k}}V}\left|\sum_{ai\bm{ k }}\langle i\bm{k}|\hat{\bm r} |a\bm{k}\rangle X_{ai\bm{k}}^{S}\right|^{2}\delta(\omega-\Omega_{S})\, ,
	\label{eq:abs_len}
\end{align}
\end{subequations}
where the velocity operator is defined as $\hat{\bm v}\equiv i[\hat H,\hat{\bm r}]$, $V$ is the volume of the unit cell, and $E_{a\bm k}$ denotes the Kohn-Sham eigenvalue associated with the single-particle state $|a\rangle$. Eq.~\eqref{eq:abs_vel} presents the expression in the velocity form and Eq.~\eqref{eq:abs_len} in length form. We will compare these two forms in Sec.~\ref{sec: basis} and derive them in Appendix ~\ref{sec:spectrum derivation}. It is worth noting that the transition amplitude $\langle i\bm{k}|\hat{\bm v} |a\bm{k}\rangle$ is non-zero only in the spin-singlet channel; consequently, only this channel contributes to the BSE spectrum. Accordingly, the BSE eigenvector should be divided by $\sqrt{2}$ to restore the correct normalization in the complete spin space.

For the full BSE case, the $Y$ component also needs to be included as follows:
\begin{equation}
	\begin{aligned}
	\varepsilon_2(\omega) =& \sum_{S}\frac{4\pi^{2}}{3N_{\bm{k}}V}\Bigg|\sum_{ai\bm{ k }}\frac{\langle i\bm{k}|\hat{\bm v} |a\bm{k}\rangle}{E_{a\bm{k}}-E_{i\bm{k}}} X_{ai\bm{k}}^{S} \\
	& + \frac{\langle a\bm{k}|\hat{\bm v} |i\bm{k}\rangle}{E_{i\bm{k}}-E_{a\bm{k}}} Y_{ai\bm{k}}^{S} \Bigg|^{2}\delta(\omega-\Omega_{S})\, .
	\end{aligned}
    \label{eq:full_BSE_epsilon2}
\end{equation}
Here, only the expression for the velocity formulation is presented, and the one for the length formulation is given analogously.

\section{Implementation Details}\label{sec:Details}
\subsection{LRI technique}
To construct the BSE Hamiltonian matrix, we employ the LRI technique~\cite{Lin2020LRI,Lin2021LRI,Cao2025,Ihrig2015,Levchenko2015}, whereby the product of two NAOs is expanded in terms of auxiliary basis functions (ABFs):
\begin{equation}
	\begin{aligned}
	&\phi_{s}(\bm r-\bm R_{s}-\bm{\tau}_{s})\phi_{t}(\bm r-\bm R_{t}-\bm{\tau}_{t}) \\
	=&\sum_{\mu\in S}C_{s(\bm R_{s}),t(\bm R_{t})}^{\mu(\bm R_{s})}P_{\mu}(\bm r-\bm R_{s}-\bm{\tau}_{s}) \\
	& + \sum_{\mu\in T}C_{t(\bm R_{t}),s(\bm R_{s})}^{\mu(\bm R_{t})}P_{\mu}(\bm r-\bm R_{t}-\bm{\tau}_{t})\, .
	\end{aligned}
\end{equation}
Here, $P_\mu(\bm r - \bm R_\mu - \bm \tau_\mu)$ is an ABF, $S$ and $T$ denote the atoms on which the two NAOs are centered. $C_{s(\bm R_{s}),t(\bm R_{t})}^{\mu(\bm R_{\mu})}$ is the LRI expansion coefficient, and $\mu\in\{S,T\}$ means that the ABFs are centered on the same atoms as the NAOs, i.e., either on atom $S$ or atom $T$. Owing to the translational symmetry, this coefficient depends only on the relative lattice vector $\bm R \equiv \bm R_t - \bm R_s$, which is equivalently denoted as $C_{s(\bm 0),t(\bm R)}^{\mu(\bm 0)}$ or $C_{st}^{\mu}(\bm R)$.

Define the Coulomb matrix element as
\begin{equation}
    V_{\mu\nu}(\bm{R})\equiv \iint \mathrm d\bm r\mathrm d\bm r' P_{\mu}(\bm r-\bm \tau_\mu) V(\bm r, \bm r') P_{\nu}(\bm r'-\bm R - \bm \tau_\nu)\, ,
\end{equation}
the LRI expansion coefficients can be obtained by minimizing the fitting error with the Coulomb metric \cite{Ihrig2015}, 
\begin{equation}
	\begin{aligned}
	C_{st}^{\mu}(\bm R) =& \sum_{\substack{\nu\in\{S,T\} \\ \bm R_\nu\in\{0,\bm R\}}} {\left[V^{ST}\right]}_{\mu\nu}^{-1} \iint \mathrm d\bm r\mathrm d\bm r' P_{\nu}(\bm r-\bm R_\nu-\bm \tau_\nu) \\
	&\times V(\bm r, \bm r') \phi_{s}(\bm r'-\bm \tau_s)\phi_{t}(\bm r'-\bm R -\bm \tau_t)\, ,
	\end{aligned}
\end{equation}
where $V^{ST}$ represents the sub-block of the Coulomb matrix, with its elements restricted to the ABFs centered on atom $S$ or $T$. The Fourier transform of the expansion coefficients is defined with respect to this single $\bm R$ variable
\begin{equation}
	C_{st}^{\mu}(\bm{ k })\equiv\sum_{R}C_{st}^{\mu}(\bm R)\mathrm{e}^{i\bm{ k }\cdot \bm{ R }}\, .
\end{equation}
For Bloch-summed NAO basis functions $\phi_{s}^{\bm k}(\bm r) = \sum_{\bm R} \phi_s(\bm r - \bm R - \bm \tau_s) e^{i\bm k \cdot \bm R}$, the product of two such $\bm k$-dependent basis functions can be expanded in terms of ABFs as
\begin{equation}
	\phi_{s}^{\bm{k}_2*}(\bm r)\phi_{t}^{\bm{k}_1}(\bm r) = \left[\sum_{\mu\in S}C_{st}^{\mu}(\bm k_1)+\sum_{\mu\in T}C_{ts}^{\mu}(-\bm k_2) \right]P_{\mu}^{\bm k_1-\bm k_2}(\bm r)\, .
\end{equation}
Therefore, we define the LRI expansion coefficients in the Kohn-Sham orbital basis as
\begin{equation}
	\widetilde C^{\mu}_{j\bm k_2,i\bm k_1} = \sum_{st} c^{\bm k_2*}_{sj} c_{ti}^{\bm k_1} [C^{\mu}_{st}(\bm k_1) + C^{\mu}_{ts}(-\bm k_2)]\, .
\end{equation}
With these definitions, the BSE Hamiltonian matrix elements can be calculated using the following formulae,
\begin{subequations}
	\begin{align}
	(j\bm{k}_2^*,i\bm{k}_1 &| W | a\bm{k}_1^*,b\bm{k}_2) \notag \\
	=& \frac{1}{N_{\bm{k}}}\sum_{\mu\nu}
		\widetilde C^{\mu}_{j\bm{k}_2, i\bm{k}_1}
		\sum_{\bm{R}} W_{\mu\nu}(\bm{R})
		\mathrm{e}^{\,i(\bm{k}_2-\bm{k}_1)\cdot\bm{R}}
		\widetilde C^{\nu}_{ a\bm{k}_1, b\bm{k}_2} \label{eq:W_LRI} \\
	(a\bm{k}_1^*,i\bm{k}_1 &| V | j\bm{k}_2^*,b\bm{k}_2) \notag \\
	=& \frac{1}{N_{\bm{k}}}\sum_{\mu\nu}
		\widetilde C^{\mu}_{a\bm{k}_1, i\bm{k}_1}
		V_{\mu\nu}(\bm{0})\,
		\widetilde C^{\nu}_{j\bm{k}_2, b\bm{k}_2}\, .
	\end{align}
\end{subequations}
where 
\begin{equation}
    W_{\mu\nu}(\bm{R})\equiv \iint \mathrm d\bm r\mathrm d\bm r' P_{\mu}(\bm r-\bm \tau_\mu) W(\bm r, \bm r', \omega=0) P_{\nu}(\bm r'-\bm R - \bm \tau_\nu)
\end{equation}
is the screened interaction with the real-space lattice vector label, and 
\begin{equation}
    V_{\mu\nu}(\bm{0})\equiv \sum_{R} \bar V_{\mu\nu}(\bm{R})
\end{equation}
is the truncated bare Coulomb interaction at the $\Gamma$ point ($\bm q =\bm 0$). We point out that the bare Coulomb interaction $\bar V$ used here is in fact truncated to tackle the divergence at the $\Gamma$ point, as will be discussed below. The above formulation essentially follows those presented in Ref.~\cite{Zhou2025}, while the only difference is that NAOs here are pseudized basis functions compatible with the norm-conserving pseudopotentials and the integrals are evaluated using the LRI infrastructures implemented in the ABACUS code \cite{Lin2020LRI,Lin2021LRI,gong2026}. 

One technical challenge is that the bare Coulomb potential displays a $1/q^2$ divergence as $\bm q \to \bm 0$ in three-dimensional systems. In the representation of plane-wave basis functions $e^{i\bm G \cdot \bm r}$, the bare Coulomb matrix has a well-known form given by $V_{\bm G, \bm G'} = 4\pi \delta(\bm G, \bm G')/|\bm q+\bm G|^2$ and diverges at $\bm G = \bm G' = \bm q = \bm 0$. In the representation of atom-centered ABF, this divergence manifests in the matrix elements of two nodeless s-type functions, displaying a $1/q^2$ behavior, and in those between one nodeless s-type and one nodeless p-type function \cite{Ren2021}, displaying a $1/q$ behavior.  These two divergent terms correspond to the ``head'' and ``wing'' terms, respectively, in the plane-wave representation, requiring special treatment to ensure numerical stability and accuracy.

To address this problem, we adopt two methods: First, when calculating the symmetrized dielectric function in the ABF representation
\begin{equation}\label{eq:dielectric}
\tilde\varepsilon_{\mu\nu}(\bm q, i\omega) = \delta_{\mu\nu} - \sum_{\alpha\beta} V_{\mu\alpha}^{1/2}(\bm q) \chi^0_{\alpha\beta}(\bm q,i\omega)V_{\beta\nu}^{1/2}(\bm q)\, ,
\end{equation}
we will treat ``head'' and ``wing'' terms with the analytical form in the representation of eigenvectors of the bare Coulomb matrix, regarding the eigenvector with the maximum eigenvalue as the $\bm G=0$ component in the plane-wave representation \cite{Ren2021,Zhang2026,gong2026}. Second, when calculating the screened Coulomb matrix, we use
\begin{equation}\label{eq:screened}
	W_{\mu\nu}(\bm q,i\omega) = \sum_{\alpha\beta}\bar V_{\mu\alpha}^{1/2}(\bm q) \tilde\varepsilon_{\alpha\beta}^{-1}(\bm q, i\omega) \bar V_{\beta\nu}^{1/2}(\bm q)\, ,
\end{equation}
where $\bar V$ is a simplified smooth form of the truncated Coulomb operator introduced by Spencer and Alavi \cite{Spencer2008,Ren2021}, which suppresses the long-range tail of the Coulomb potential. Further details of the truncation scheme are provided in Sec.~S2 of the Supplemental Material.


\subsection{Workflow of dual-$\bm k$-mesh strategy}\label{sec:workflow}

A key feature of our approach is that the screened Coulomb interaction $W$ is evaluated on a relatively coarse $\bm k$-mesh and then Fourier-interpolated to a dense $\bm k$-mesh for constructing the BSE Hamiltonian. In practice, $W_{\mu\nu}(\bm k)$ is first computed on the coarse $\bm k$-mesh and then inverse-Fourier transformed to $W_{\mu\nu}(\bm R)$, which resides on a correspondingly small Born--von~K\'arm\'an supercell. The accuracy of this finite-supercell representation depends on the real-space range of $W$ and can be checked by refining the coarse $\bm k$-mesh. For Si with the smooth cutoff introduced above, the BSE spectra obtained from $7^3$ and $8^3$ coarse $\bm k$-meshes differ from the $9^3$ result by only 2.6\% and 2.3\% (relative $L_2$ norm) at a fixed $14^3$ fine $\bm k$-mesh (Sec.~S2 of the Supplemental Material). The computation of the dielectric function and screened interaction via Eqs.~\eqref{eq:dielectric} and~\eqref{eq:screened} dominates the overall cost and therefore benefits most from the reduced $\bm k$-point sampling. Finally, $W_{\mu\nu}(\bm R)$ is Fourier transformed to $W_{\mu\nu}(\bm q)$ on a denser target $\bm k$-mesh for assembling and diagonalizing the BSE Hamiltonian.

Because the interpolation in Eq.~\eqref{eq:W_LRI} is applied to the screened interaction $W$ itself, rather than to the final spectrum, the BSE Hamiltonian on the dense $\bm k$-mesh is constructed from physically meaningful matrix elements at every $\bm k$-point. The dual-$\bm k$-mesh strategy thus provides a practical bridge between affordable coarse $\bm k$-mesh for $GW$ calculations and converged dense $\bm k$-mesh for excitonic spectra. 

\begin{figure}[htbp]
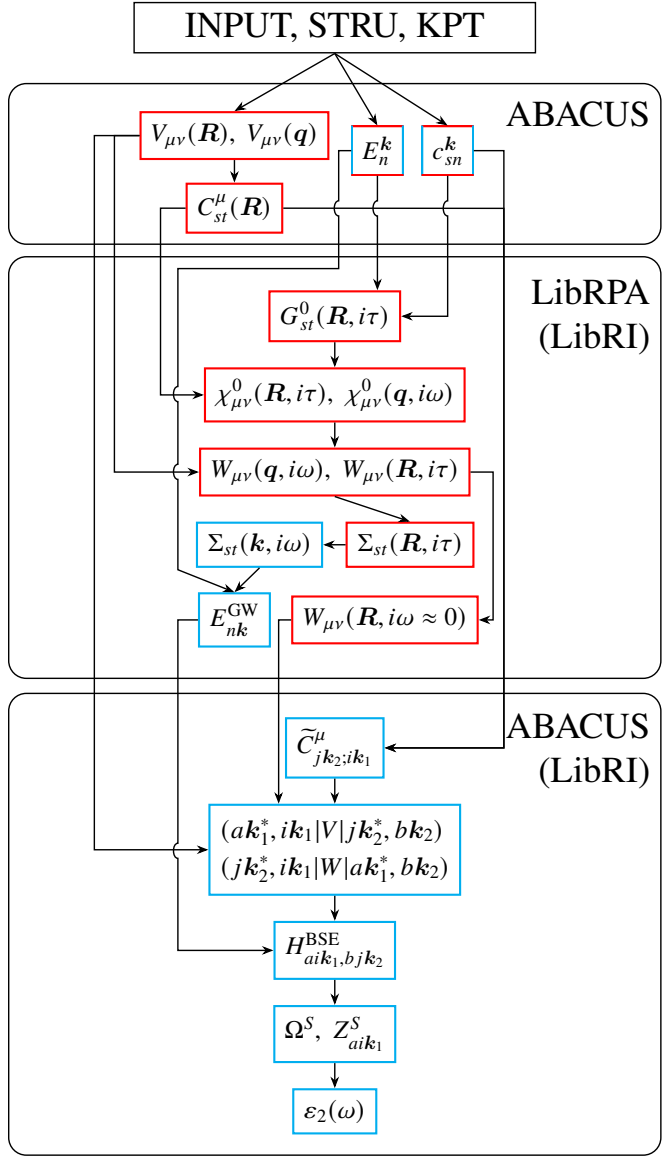

	\centering
	\includestandalone[mode=tex,width=\linewidth]{fig/workflow}
	\caption{Workflow of the BSE implementation with dual-$\bm k$-mesh strategy. Red frame indicates calculation on the coarse $\bm k$-mesh; blue frame indicates calculation on the dense $\bm k$-mesh. The BSE module resides in ABACUS and relies on quasiparticle energies and screened interactions from the $G^0W^0$ implementation in LibRPA\cite{librpa-repo}. The LRI-related tensor contractions are handled by LibRI\cite{libri-repo}, an efficient parallel library based on hybrid MPI/OpenMP parallelization.}
	\label{fig:workflow}
\end{figure}

The overall dual-$\bm k$-mesh workflow is illustrated in Fig.~\ref{fig:workflow}. The key computational steps are as follows:
\begin{enumerate}
	\item Perform the DFT calculation within ABACUS (upper panel in Fig.~\ref{fig:workflow}) to obtain the Kohn--Sham energies $E_n^{\bm k}$, eigenvectors $c_{sn}^{\bm k}$, LRI expansion coefficients $C_{st}^\mu(\bm R)$, and the Coulomb matrix $V_{\mu\nu}(\bm q)$ in the ABF basis. The Coulomb matrix is first evaluated in real space as $V_{\mu\nu}(\bm R)$ for atom pairs within a cutoff set by the input parameter \texttt{exx\_rmesh\_times} times the NAO radius \texttt{rcut}, then folded into a Born--von~K\'arm\'an supercell and Fourier-transformed to $V_{\mu\nu}(\bm q)$ on the coarse $\bm k$-mesh. The coefficients $C_{st}^\mu(\bm R)$ are truncated beyond twice the NAO radius \texttt{rcut}. In this step, $V_{\mu\nu}(\bm R)$ and $C_{st}^\mu(\bm R)$ are evaluated on a relatively small Born--von~K\'arm\'an supercell ($\bm R$ mesh) that is nevertheless sufficient to accommodate the spatial extent of the NAOs and ABFs, while $E_n^{\bm k}$ and $c_{sn}^{\bm k}$ are evaluated on both the coarse and dense $\bm k$-meshes, as required for the $G^0W^0$ and BSE calculations.
	\item Perform the $G^0W^0$ calculation within LibRPA (middle panel in Fig.~\ref{fig:workflow}). Starting from the non-interacting Green's function $G_{st}^0(\bm R, i\tau)$ in the real-space imaginary-time domain, the calculation proceeds through the non-interacting response function $\chi^0_{\mu\nu}$ and the screened Coulomb matrix $W_{\mu\nu}$ in the ABF basis, to the self-energy $\Sigma_{\text{xc},st}$ in the NAO basis. These steps involve Fourier transforms between small $\bm R$ mesh and coarse $\bm q$ mesh, as well as cosine/sine transforms between $\tau$ and $\omega$~\cite{Zhang2026}, accelerated by a minimax time-frequency grid~\cite{AziziM2023,Kaltak2014,Liu2016}. Once $\Sigma_{\text{xc},st}(\bm R)$ is prepared, we Fourier transform it to a dense $\bm k$-mesh, and calculate the relevant $G^0W^0$ quasiparticle energies $E_{n\bm k}^{\text{GW}}$. The static screened Coulomb matrices $W_{\mu\nu}(\bm R, i\omega \approx 0)$ in small $\bm R$ mesh at the lowest minimax-grid frequency are also prepared during this step.
	\item Perform the BSE calculation within ABACUS (lower panel in Fig.~\ref{fig:workflow}). From this step onward, all computations will be performed on the dense $\bm k$-mesh. The LRI expansion coefficients $C_{st}^\mu(\bm R)$ are transformed from the NAO basis to the Kohn--Sham orbital basis to obtain $\widetilde C^{\mu}_{j\bm k_2,i\bm k_1}$. Then, we construct the BSE Hamiltonian matrix via Eq.~\eqref{eq:Hamiltonian} using the $G^0W^0$ quasiparticle energies and the interaction kernel of Eq.~\eqref{eq:BSEkernel}.
	\item Diagonalize the BSE Hamiltonian $H^\text{BSE}$ to obtain the exciton eigenvalues and eigenvectors. The absorption spectrum can be computed by Eq.~\eqref{eq:abs} for the TDA case or Eq.~\eqref{eq:full_BSE_epsilon2} for the full case.
\end{enumerate}

\section{Results and Discussion}\label{sec:Results}
In this section, we present a systematic assessment of the BSE implementation in ABACUS+LibRPA along two directions. First, we examine the convergence behavior with respect to the key computational parameters---the NAO basis set, the auxiliary basis set, and the $\bm k$-point sampling---and discuss strategies for accelerating convergence. Second, we benchmark the accuracy of the implementation against established codes for molecular systems (Thiel's set) and periodic solids (Si, MgO), and analyze exciton binding energies for a set of representative materials. Unless otherwise specified, the ABACUS calculations in this work use SG15 optimized norm-conserving Vanderbilt pseudopotentials~\cite{Schlipf2015SG15}.

\subsection{Convergence with various parameters}
\subsubsection{Assessing NAO basis set convergence via velocity--length equivalence}\label{sec: basis}
\begin{figure*}[htbp]
	\centering
	\includegraphics[width=0.96\textwidth]{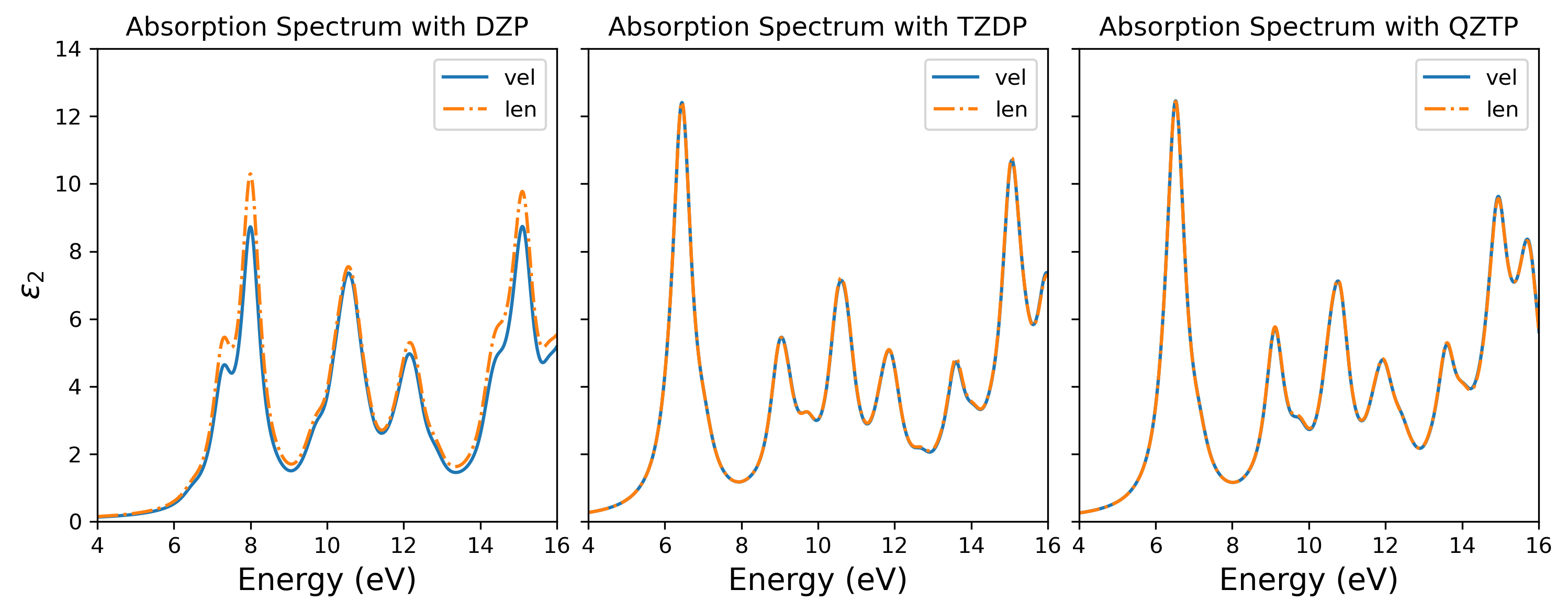}
	\caption{BSE absorption spectra of benzene computed with three NAO basis sets (DZP, TZDP, and QZTP), shown in both the velocity and length formulations. Lorentzian broadening is set to $\eta = 0.3$ eV. As the basis set is enlarged from DZP to QZTP, the discrepancy between the two formulations is systematically reduced, indicating that their residual difference mainly arises from basis set incompleteness.}
	\label{fig:ben}
\end{figure*}

In a complete basis set, the velocity and length formulations of the BSE spectra in Eq.~\eqref{eq:abs} are formally equivalent for bound systems, see Appendix \ref{sec:spectrum derivation}. However, as the basis set is incomplete in practice, the two formulations may yield different results. This behavior originates from the fact that, in our implementation, the velocity and position matrix elements are evaluated on the real-space grid, whereas the Kohn--Sham eigenvalue problem is solved in a finite NAO basis. Consequently, the eigenvalue relation $\hat H|\psi\rangle = E|\psi\rangle$, which holds in the NAO basis, may not be strictly satisfied in the real-space grid representation; it becomes exact only in the limit of a complete basis set. This type of error is commonly known as the basis set incompleteness error (BSIE)~\cite{Susi2019,Peng2023,Peng2025}. In this work, we leverage the discrepancy between the length and velocity formulations as an internal diagnostic of BSIE, and assess the convergence of BSE calculations with respect to NAO basis sets, using benzene as a test system.

Fig.~\ref{fig:ben} shows the BSE absorption spectra computed with three NAO basis sets~\cite{Chen2010,Lin2021NAO}---DZP, TZDP, and QZTP---in both formulations. For carbon, these correspond to $2s2p1d$, $3s3p2d$, and $4s4p3d$ valence-plus-polarization basis functions, respectively. The systematic convergence of the two formulations with increasing basis size confirms that their residual difference indeed arises from basis set incompleteness. Since the discrepancy at the TZDP level is already rather small, we adopt TZDP for the subsequent benchmark calculations.

For extended systems, we find that only the velocity formulation remains valid, whereas the length formulation encounters a fundamental difficulty: the position coordinate operator compromises the Hermiticity of the Hamiltonian operator, as discussed in detail in Appendix \ref{sec:length form}.

\subsubsection{Auxiliary basis set convergence}
As discussed in Sec.~\ref{sec:Details}, the present BSE implementation employs the localized resolution-of-identity (LRI) technique, in which products of two NAOs are expanded in terms of ABFs. The accuracy of the LRI approximation, and consequently of the BSE results, depends on the quality and completeness of the ABFs. In this subsection, we assess the convergence of the optical absorption spectra with respect to the ABFs by varying the threshold used in the principal component analysis (PCA) compression step in the generation of the ABFs. The PCA step linearly recombines the original ABFs (``on-site'' products of NAOs) and discards those whose variance falls below the chosen threshold~\cite{Lin2020LRI}, thereby controlling the final number of ABFs. 

\begin{figure}[htbp]
	\centering
	\includegraphics[width=0.46\textwidth]{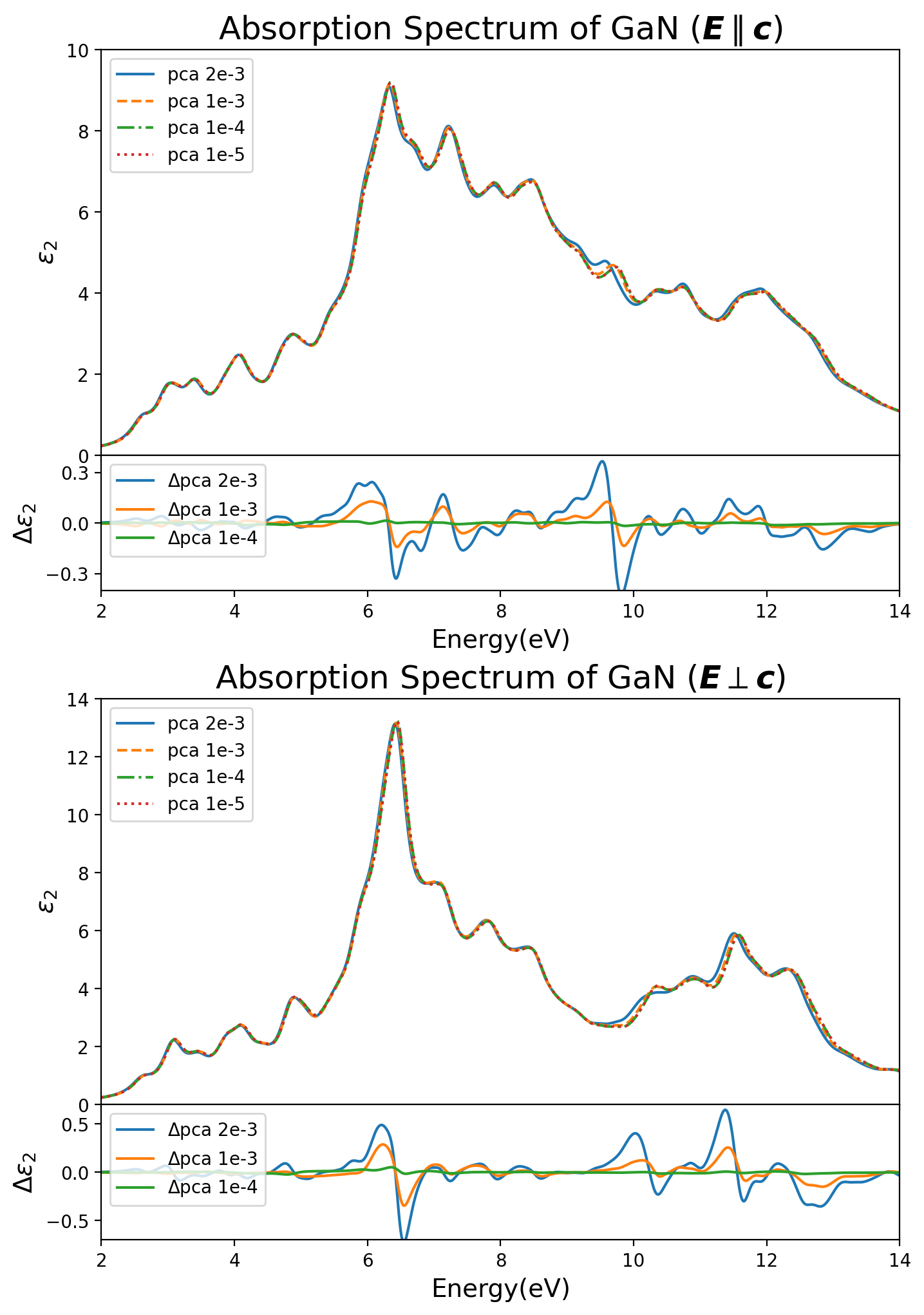}
	\caption{Convergence of the BSE absorption spectra of GaN with respect to the auxiliary basis size. All calculations use 6 occupied and 8 virtual orbitals, $15\times15\times10$ $\Gamma$-centered $\bm k$-point mesh, with Lorentzian broadening $\eta = 0.2$ eV. The PCA threshold controls the number of retained ABFs by eliminating those with variance below the threshold, resulting in 988, 1178, 1704, and 2348 functions for thresholds of $2\times10^{-3}$, $10^{-3}$, $10^{-4}$, and $10^{-5}$, respectively.}
	\label{fig:GaN}
\end{figure}

Given that heavier elements exhibit increased sensitivity to LRI errors, we choose GaN (wurtzite structure, mp-804 \cite{Jain2013MP}) as our test system. Since the LRI-based $G^0W^0$ self-energy calculations demand highly complete ABFs, we augment the original NAO basis set with additional high angular momentum functions, referred to as the \textit{for\_aux} basis in Refs.~\cite{Ihrig2015,Ren2021}; in this case, one $f$-orbital and one $g$-orbital (1f1g) are added per atom. These augmented functions are used exclusively in the construction of the ABFs and do not participate in the preceding Kohn--Sham DFT calculations. This augmentation, while necessary for completeness, further increases the computational cost. To balance accuracy and efficiency, we employ PCA independently for each element to linearly recombine the initial ABF set $\{|P_\mu\rangle\}$ into a new ABF set $\{|P_\gamma '\rangle\}$ that maximizes the variance of the overlap coefficients $\langle P_\gamma'|\phi_s\phi_t\rangle$; this process is equivalent to solving an eigen-problem of the covariance matrix $\sum_{st}\langle P_\mu|\phi_s\phi_t\rangle \langle \phi_s\phi_t |P_\nu\rangle $~\cite{Lin2020LRI}. By setting an appropriate threshold, we retain only those ABFs whose contributions to the variance exceed the threshold value, which means they have the most significant impact on the electronic structure. The resulting numbers of ABFs in one unit cell are 988, 1178, 1704, and 2348 for PCA thresholds of $2\times10^{-3}$, $10^{-3}$, $10^{-4}$, and $10^{-5}$, respectively.

Fig.~\ref{fig:GaN} shows the absorption spectra for two polarization directions, $\bm E \parallel \bm c$ and $\bm E \perp \bm c$. The spectra computed with different PCA thresholds are nearly identical for both polarizations, indicating that the error introduced by the LRI approximation is negligible across the tested range of auxiliary basis sizes. Additionally, we have plotted the spectral differences relative to the PCA threshold of $10^{-5}$ in the inset of the figure, where the maximum deviation remains below 0.3 a.u., confirming the excellent agreement among the spectra. This observation is consistent with the findings of Zhou \textit{et al.} \cite{Zhou2025}, who reported that the sensitivity of BSE calculations to the auxiliary basis sets is less pronounced than that of the $G^0W^0$ quasiparticle calculations. The physical reason for this reduced sensitivity is that the screened interaction entering the BSE kernel primarily involves electron-hole pairs near the Fermi level, whereas the $G^0W^0$ self-energy requires the screened interaction for high-lying orbitals as well, necessitating a more complete auxiliary basis for converged quasiparticle energies.

The dominant absorption peak is observed at approximately 6.4~eV, with the $\bm E \perp \bm c$ polarization showing a significantly higher intensity than the $\bm E \parallel \bm c$ polarization, reflecting the optical anisotropy of the wurtzite crystal structure. These features are consistent with the work of Laskowski \textit{et al.} \cite{Laskowski2005}. 

\subsubsection{$\bm k$-convergence and $\bm k$-offset} \label{sec:k}

One of the most important and computationally demanding aspects of BSE calculations for periodic systems is achieving convergence with respect to the BZ sampling.  The $\bm k$-point convergence of the BSE absorption spectrum is considerably slower than that of the underlying $G^0W^0$ quasiparticle energies, as the BSE spectrum exhibits a stronger dependence on wavefunction sampling for computing transition dipole moments. For Si and MgO, the $G^0W^0$ band gap is already converged on a $7\times7\times7$ mesh~\cite{Ren2021}, whereas the corresponding BSE spectrum still exhibits significant changes in peak positions and intensities at this mesh density.

\begin{figure}[htbp]
	\centering
	\includegraphics[width=0.46\textwidth]{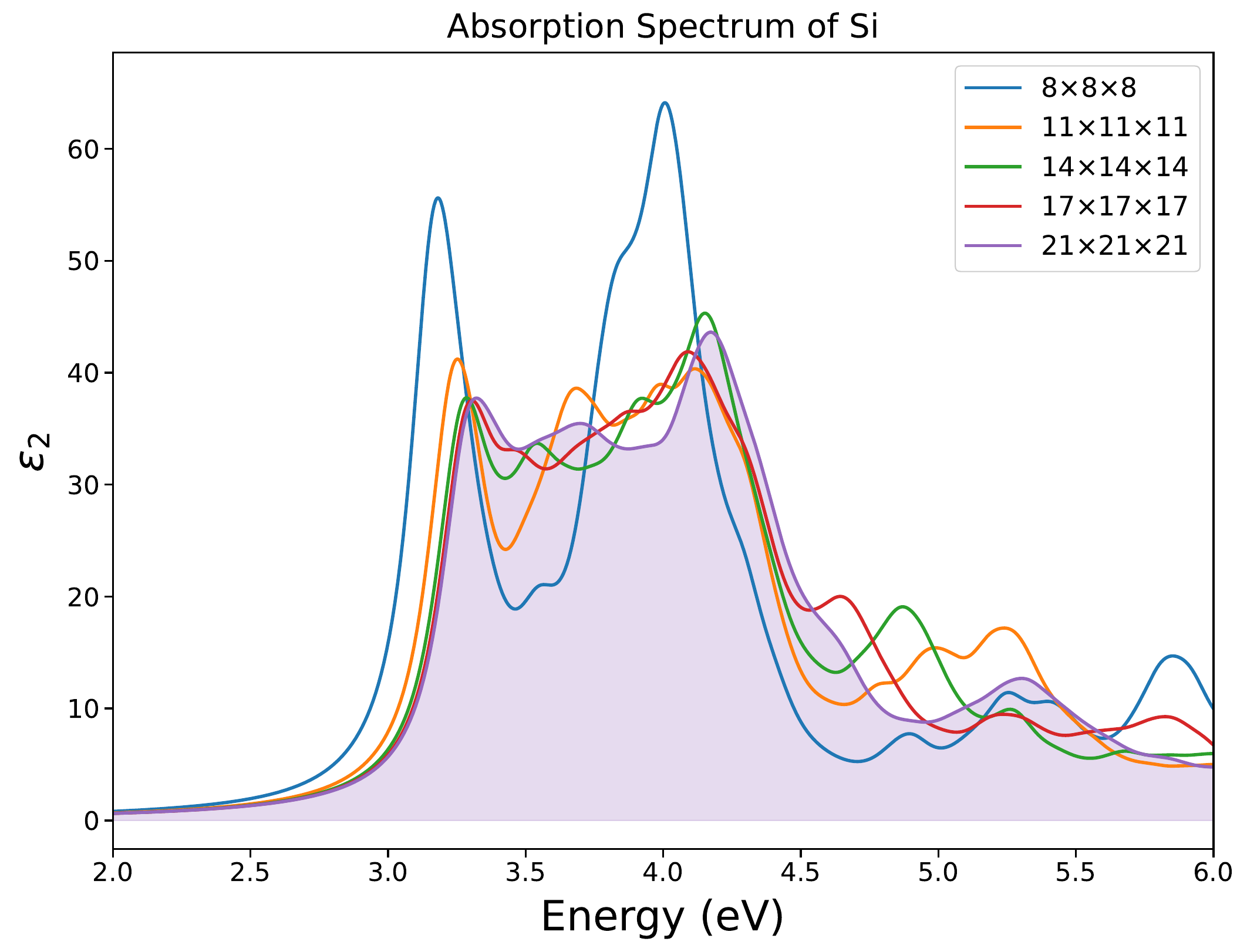}
	\caption{Convergence of the absorption spectrum with $\bm k$-point sampling. All calculations use 4 occupied and 4 virtual orbitals, with Lorentzian broadening $\eta = 0.1$ eV. The $G^0W^0$ calculation uses a $7\times7\times7$ $\Gamma$-centered $\bm k$-point mesh, and Fourier interpolation is used to obtain the BSE kernel on the target $\bm k$-point meshes.}
	\label{fig:k-conv}
\end{figure}

We examine the dense $\bm k$-mesh convergence behavior in Fig.~\ref{fig:k-conv}. All calculations use 4 occupied and 4 virtual orbitals with Lorentzian broadening of $\eta = 0.1$~eV. As the BZ sampling is refined, the spectrum generally evolves toward its converged form, but the convergence is not uniform across all energies, especially in the region from 4.5 to 5.5~eV, where the peak positions and intensities exhibit significant changes even at the $21\times21\times21$ mesh. BSE wall times and memory requirements for different dense $\bm k$-meshes are reported in Sec.~S1 of the Supplemental Material, while coarse $\bm k$-mesh convergence is examined in Sec.~S2.

To mitigate these sampling effects, shifted $\bm k$-meshes can be used in two ways. Benedict and Shirley~\cite{Benedict1999} employed shifted meshes in BSE calculations for GaN and CaF$_2$. Vorwerk~\textit{et al.}~\cite{Vorwerk2019} subsequently used this approach for LiF. A suitable shift can reduce symmetry redundancy and improve Brillouin-zone sampling at a fixed mesh density. However, the choice of shift is system dependent. An alternative is to average spectra from independent calculations on meshes with different offsets~\cite{Paier2008,Gillet2013,Sander2015}. This approach reduces sensitivity to any individual shift and can yield smoother spectra. Here, we construct offsets from a symmetry-reduced $3\times3\times3$ grid. The four representative offsets are $(0,0,0)$, $(1/3,0,0)$, $(1/3,1/3,0)$, and $(2/3,1/3,0)$, with respective weights $1/27$, $8/27$, $6/27$, and $12/27$.

\begin{figure}[htbp]
	\centering
	\includegraphics[width=0.46\textwidth]{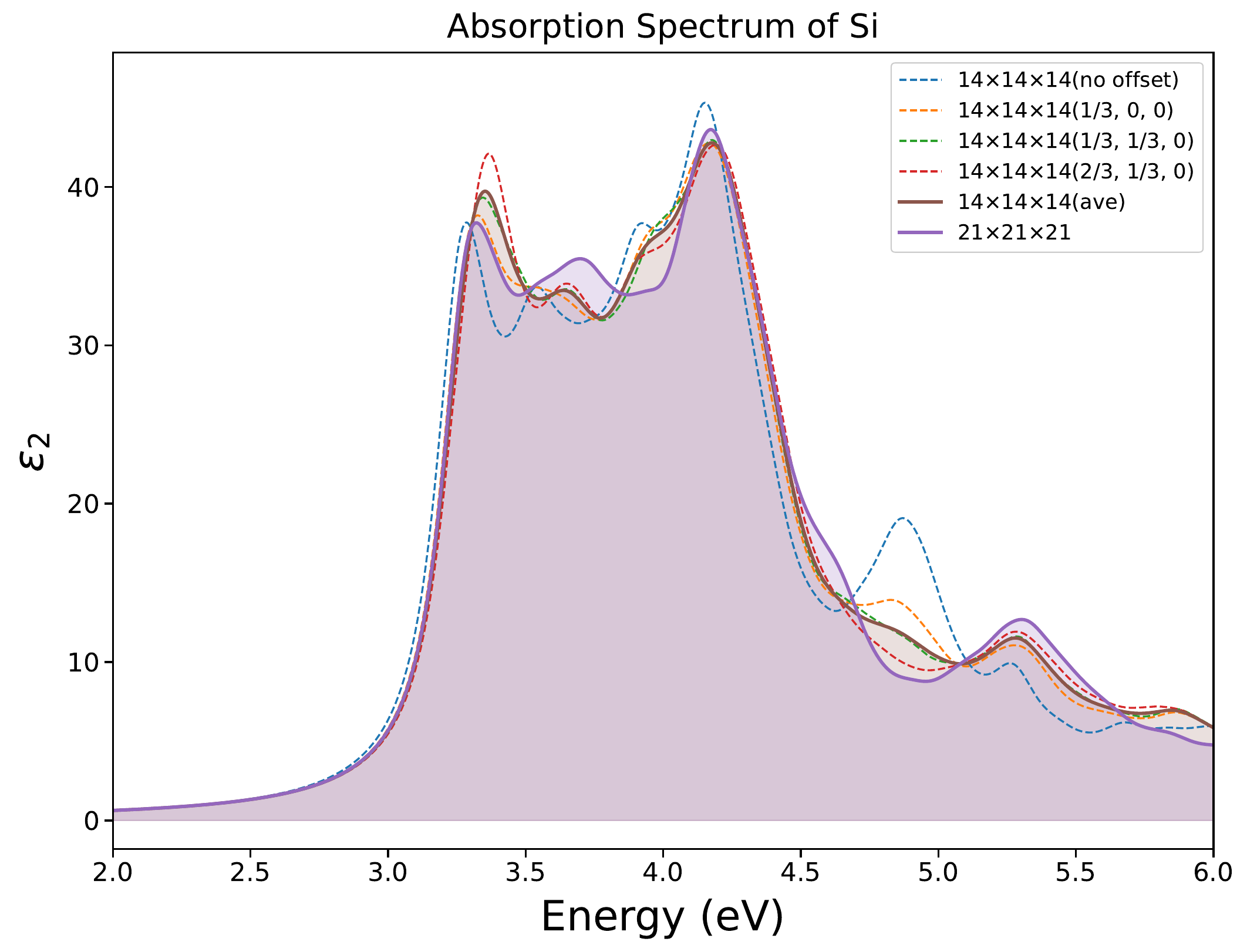}
	\caption{Approaching the dense $\bm k$-mesh spectrum via $\bm k$-point offset averaging. Computational settings are the same as Fig.~\ref{fig:k-conv}. }
	\label{fig:k-off}
\end{figure}

As shown in Fig.~\ref{fig:k-off}, the unshifted ($\bm{k}$-offset $=0$) $14\times14\times14$ mesh exhibits an anomalously large peak around 4.87~eV compared with the $21\times21\times21$ mesh. This artifact arises from the disproportionate contribution of high-symmetry $\bm k$-points when the mesh does not adequately sample the transition density in the relevant region of the BZ~\cite{Sander2015}. Each nonzero offset shown here breaks the high symmetry of the unshifted mesh and yields a spectrum much closer to the $21\times21\times21$ result. Averaging over the four symmetry-reduced offsets with appropriate weights likewise suppresses the anomalous peak and produces a spectrum close to this denser-mesh result. The purpose of offset averaging is to reduce the uncertainty associated with the particular choice of offset, rather than to ensure a closer match than every individual shifted spectrum. Since the $21\times21\times21$ result is itself not fully converged, it serves only as a denser-mesh reference.

We note that Sander \textit{et al.}~\cite{Sander2015} report a spectrum in closer agreement with experiment \cite{Aspnes1983} as their calculation is based on sc-$GW_0$. It is well established that $G^0W^0$ quasiparticle band gaps systematically underestimate experimental values (1.16 eV vs.\ 1.23 eV in the present case), whereas sc-$GW_0$ can accurately reproduce experimental band gaps. It is also worth noting that offset averaging does not eliminate the need for sufficiently dense underlying meshes; rather, it provides a practical route to obtaining smoother, more representative spectra at a moderate computational cost.

\subsection{Benchmark with other packages and experiment}

\subsubsection{Molecular systems: Thiel's set}
To assess the accuracy of the present implementation, we benchmark against Thiel's set~\cite{Schreiber2008}, which contains 28 molecules. For this test set, ABACUS+LibRPA results are compared directly with FHI-aims~\cite{Liu2020} for both spin channels (singlet and triplet) and for both BSE variants (TDA and full BSE). To minimize errors associated with NAO basis differences, we use the same Tier2 basis set in both codes. Furthermore, the ground-state PBE calculation and the construction of LRI-related tensors are performed entirely within FHI-aims, thereby ensuring identical Kohn--Sham orbitals and eigenvalues as the starting point and avoiding discrepancies from different numerical integration schemes. Subsequent dielectric-function-related calculations ($G^0W^0$ and BSE) are then performed within each code's native workflow. All calculations include all occupied orbitals and up to 300 virtual orbitals. For each molecule and each channel/variant, the lowest five excitations are considered in the analysis, referred to hereafter as the first to fifth excitation states. Detailed data are provided in the supplementary file `Thiel-test.xlsx', together with the corresponding PBE and $G^0W^0$ HOMO-LUMO gaps.

As summarized in Table~\ref{tab:thiel_accuracy}, ABACUS+LibRPA shows good agreement with FHI-aims across all four categories. For singlet excitations, the mean absolute deviation (MAD) is 32.0~meV in TDA and 35.1~meV in full BSE; for triplet excitations, the MAD is 43.0~meV (TDA) and 43.9~meV (full BSE). For comparison, the $G^0W^0$ HOMO-LUMO gap difference yields an MAD of 23.7~meV, indicating that the quasiparticle-level agreement is excellent. The larger BSE-level discrepancies therefore originate from differences in the BSE kernel construction between the two codes' native workflows. Since the Kohn--Sham starting point and LRI tensors are identical, the one implementation detail that can be firmly identified as differing is the frequency treatment of the screened Coulomb interaction $W$: LibRPA employs a minimax time-frequency grid, while FHI-aims uses a modified Gauss--Legendre grid~\cite{Liu2020,Ren2021}.

The corresponding mean signed deviations (MSD) are small: $-$11.7 and $-$7.0~meV for singlet TDA and full BSE, and $+$4.3 and $+$3.7~meV for triplet TDA and full BSE, respectively. The distribution of discrepancies is also well controlled: over 85\% of the data points lie within 100~meV (95.7\% for singlet TDA, 93.6\% for singlet full BSE, 85.7\% for triplet TDA, and 88.6\% for triplet full BSE). The maximum absolute deviations (MaxAD) are 214.6, 286.4, 290.1, and 294.1~meV for singlet TDA, singlet full BSE, triplet TDA, and triplet full BSE, respectively, with the largest values concentrated in a small number of molecules.

The most systematic pattern in the data is the dependence of the discrepancy on conjugation length in the linear polyene series: ethene (C$_2$), \emph{E}-butadiene (C$_4$), all-\emph{E}-hexatriene (C$_6$), and all-\emph{E}-octatetraene (C$_8$). Ethene, with a single isolated double bond, shows near-perfect agreement between the two codes (MAD $<$10~meV across all four channels). As conjugation extends, the MAD grows: for the triplet full BSE channel, the per-molecule MAD increases from 7.6~meV (ethene) to 79.5~meV (butadiene), 134.6~meV (hexatriene), and 208.0~meV (octatetraene). A similar monotonic trend is observed in triplet TDA (5.8 $\to$ 87.4 $\to$ 138.9 $\to$ 175.9~meV). The singlet channels show a weaker and less monotonic chain-length dependence, with octatetraene (MAD 179.5~meV in full BSE) clearly separating from the shorter polyenes (MAD $\sim$40--50~meV). The growth of the discrepancy with chain length may reflect the increasing spatial extent of the exciton wave function: a more delocalized exciton samples a larger region of the BSE kernel, thereby accumulating the effect of small numerical differences in the implementation over a larger effective volume.

Beyond the polyene series, the largest per-molecule deviation occurs in \emph{s}-tetrazine (molecule~18, MAD 130.1~meV). The discrepancies are concentrated in specific excited states rather than distributed uniformly across the spectrum. The lowest excitations agree within 10--20~meV, whereas deviations exceed 200~meV for certain higher-lying states. \emph{s}-Tetrazine contains $n\to\pi^*$ states involving nitrogen lone pairs, and the discrepancies may be associated with the character of particular excited states. The origin of this state dependence remains to be clarified.

Taken together, these results demonstrate that ABACUS+LibRPA maintains a consistent accuracy level from TDA to full BSE and from singlet to triplet channels. The residual discrepancies exhibit three characteristic patterns: an overall MAD of 30--50~meV for the whole Thiel's set, a systematic growth with conjugation length in linear polyenes (most pronounced in the triplet channels), and a pronounced dependence of the deviation on excitation order, with higher-lying states generally showing larger discrepancies.

\begin{table}[h]
	\caption{Statistical deviations between ABACUS+LibRPA and FHI-aims on the Thiel's set (28 molecules, 140 excitations per category). All values are in meV except the last column. }
	\begin{ruledtabular}
		\begin{tabular}{lcccc}\label{tab:thiel_accuracy}
			Case               & MSD  & MAD  & MaxAD & $|\Delta E|\le 0.1$ eV \\
			\hline
			Singlet (TDA)      & $-$11.7 & 32.0 & 214.6 & 95.7\%                 \\
			Singlet (full BSE) & $-$7.0  & 35.1 & 286.4 & 93.6\%                 \\
			Triplet (TDA)      & +4.3  & 43.0 & 290.1 & 85.7\%                 \\
			Triplet (full BSE) & +3.7  & 43.9 & 294.1 & 88.6\%                 \\
		\end{tabular}
	\end{ruledtabular}
\end{table}

\subsubsection{Periodic systems: Si and MgO}
We next benchmark the present implementation against established references for two prototypical systems: silicon (Si), a medium-gap covalent semiconductor with weakly bound excitons, and magnesium oxide (MgO), a wide-gap ionic insulator with strongly bound excitons.

\begin{figure}[htbp]
	\centering
	\includegraphics[width=0.46\textwidth]{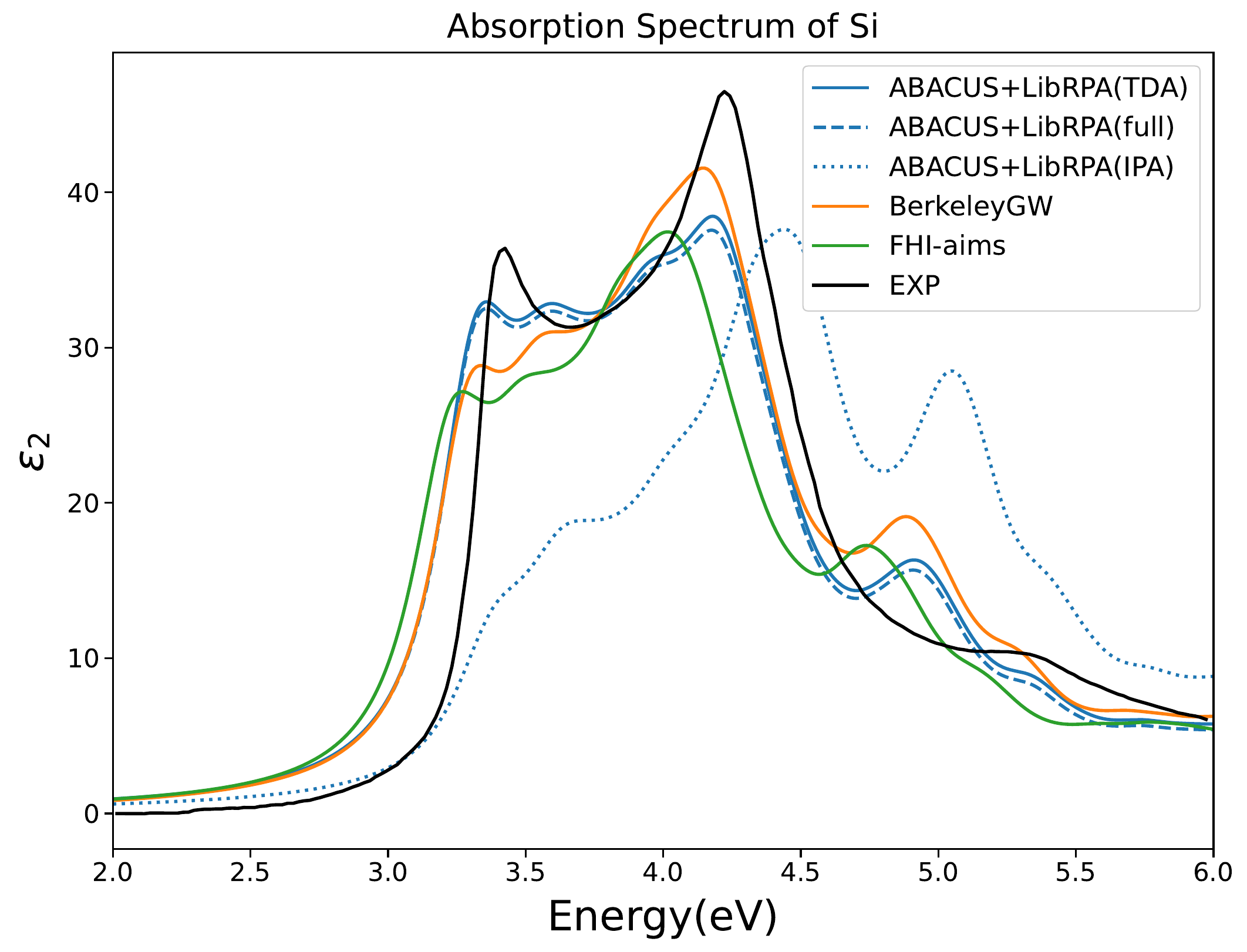}
	\caption{BSE absorption spectra for Si with a $14\times14\times14$ $\Gamma$-centered $\bm k$-point mesh. All calculations use 4 occupied and 4 virtual orbitals, with Lorentzian broadening $\eta = 0.15$ eV. ABACUS+LibRPA uses SG15 pseudopotentials and the TZDP basis. BerkeleyGW and FHI-aims curves are taken from Zhou \textit{et al.}~\cite{Zhou2025}. Experimental data are from Aspnes and Studna~\cite{Aspnes1983}.}
	\label{fig:Si}
\end{figure}

Fig.~\ref{fig:Si} compares the Si absorption spectrum computed with ABACUS+LibRPA against published FHI-aims and BerkeleyGW results~\cite{Zhou2025} on the same $14\times14\times14$ $\Gamma$-centered mesh. All calculations use 4 occupied and 4 virtual orbitals with a Lorentzian broadening of $\eta = 0.15$~eV. Experimental data from Aspnes and Studna~\cite{Aspnes1983} are also shown for reference.

The overall line shape and major peak positions are in close agreement across all three codes. The agreement between ABACUS+LibRPA and FHI-aims is particularly meaningful, as both codes employ NAO basis sets with the LRI technique for constructing the screened interaction, thereby minimizing discrepancies arising from different basis representations. The comparison with BerkeleyGW~\cite{Deslippe2012}, which uses a plane-wave basis, further validates the present implementation across fundamentally different numerical frameworks. Despite the close overall agreement, minor differences in peak intensities are visible in the energy range 3.0--4.5~eV. These residual discrepancies can be attributed to several factors, including differences in the frequency treatment used to obtain the preceding $GW$ quasiparticle corrections and in the handling of the Coulomb singularity at $\bm q \to 0$.

Compared with the experimental spectrum~\cite{Aspnes1983}, all three calculations exhibit a slight redshift and somewhat broader peak structures, which is typical for BSE calculations at the $G^0W^0$ level. Sander \textit{et al.}~\cite{Sander2015} performed the calculation based on sc-$GW_0$, which yields better agreement with the experiment due to its ability to accurately reproduce the experimental band gap. The comparison with the experiment is further affected by the absence of electron–phonon coupling and temperature effects in the theoretical treatment. Nevertheless, the close agreement among the three independent codes confirms that ABACUS+LibRPA correctly reproduces the expected spectral features for a prototypical covalent semiconductor with weakly bound excitons.

The TDA and full BSE spectra for Si are almost indistinguishable in $\varepsilon_2$, as shown in Fig.~\ref{fig:Si}. For comparison, the independent-particle approximation (IPA) result substantially overestimates the absorption peak energy and fails to reproduce the spectral line shape. This overestimation directly reflects the fact that the electron-hole attraction, which is absent in IPA, lowers the excitation energy through excitonic correlation --- reinforcing the essential role of the electron--hole interaction in the optical response of Si.

\begin{figure}[htbp]
	\centering
	\includegraphics[width=0.46\textwidth]{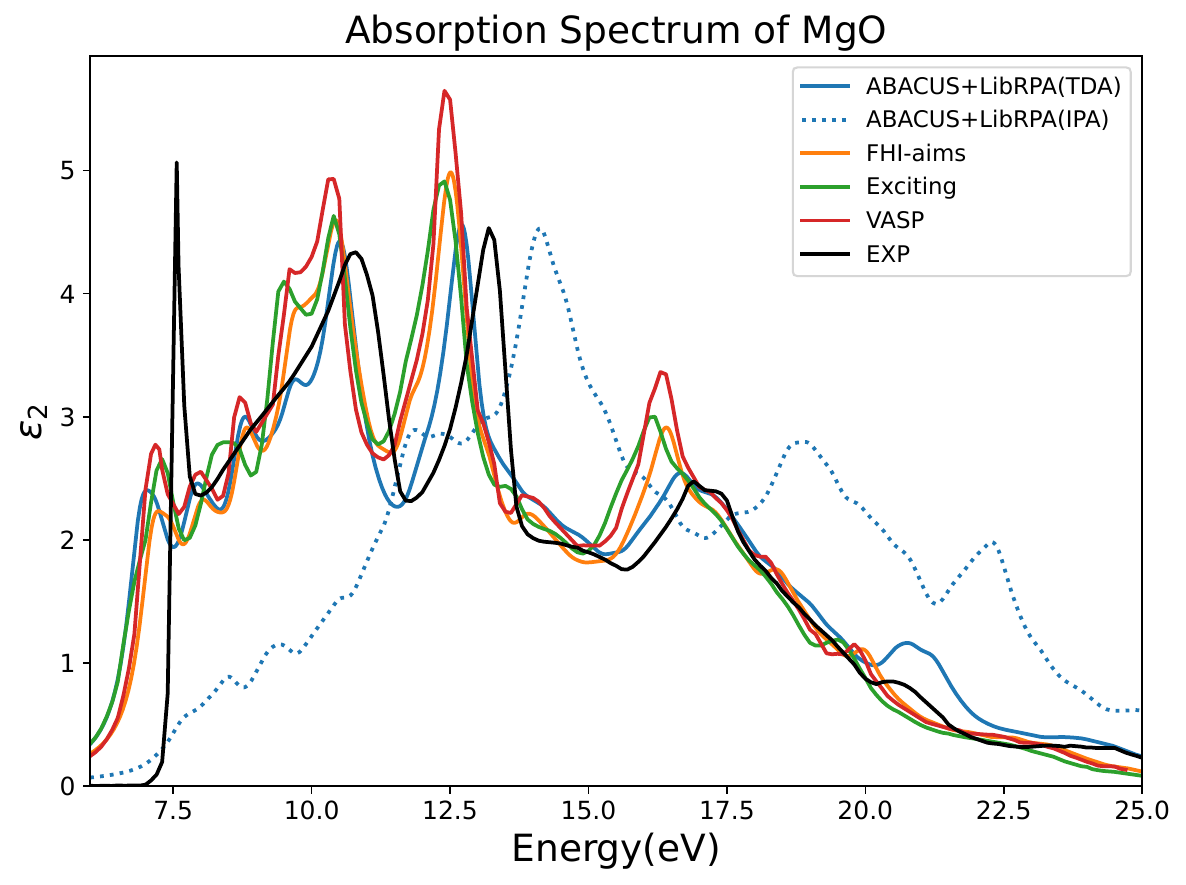}
	\caption{MgO absorption spectra with a $15\times15\times15$ $\Gamma$-centered $\bm k$-point mesh. All calculations use 4 occupied and 5 virtual orbitals, with Lorentzian broadening $\eta = 0.3$ eV. ABACUS+LibRPA uses the APNS (ABACUS Pseudopotential-Numerical atomic orbital-Square) setup~\cite{Huang2026APNS}. FHI-aims~\cite{Zhou2025} and VASP~\cite{Begum2021} references were obtained on $15\times15\times15$ $\Gamma$-centered meshes, while Exciting~\cite{Begum2021} used an $11\times11\times11$ $\Gamma$-centered mesh shifted by $(0.09,\,0.02,\,0.04)$. Experimental data are from Roessler and Walker~\cite{Roessler1967}.}
	\label{fig:MgO}
\end{figure}

Fig.~\ref{fig:MgO} shows the corresponding benchmark for MgO, compared with VASP, Exciting, and FHI-aims references from the literature, using a $15\times15\times15$ $\Gamma$-centered mesh with 4 occupied and 5 virtual orbitals and $\eta = 0.3$~eV. We note that the reference calculations employed slightly different sampling methods: FHI-aims~\cite{Zhou2025} and VASP~\cite{Begum2021} used $15\times15\times15$ $\Gamma$-centered meshes, while Exciting~\cite{Begum2021} used an $11\times11\times11$ $\Gamma$-centered mesh shifted by $(0.09,\,0.02,\,0.04)$.

Despite substantial differences in implementation---including the choice of basis set (NAO for ABACUS+LibRPA and FHI-aims, plane waves for VASP~\cite{Sander2015}, linearized augmented plane waves for Exciting~\cite{Vorwerk2019}), the treatment of core electrons, and the Brillouin zone sampling strategy---ABACUS+LibRPA reproduces the main peak structure and relative intensity distribution in good agreement with the reference data.  The dominant absorption features in the 7--14~eV range are well captured, and the overall line shape is consistent across all codes. The ABACUS+LibRPA peak intensities near 12.5 and 16.5~eV remain somewhat lower than those of the other codes. The spectrum in Fig.~\ref{fig:MgO} is obtained with the APNS (ABACUS Pseudopotential-Numerical atomic orbital-Square) pseudopotential/basis setup~\cite{Huang2026APNS}. With the SG15+TZDP setup used for the other benchmarks, the intensity around 10--11~eV is noticeably lower, and the two setups already differ at the Kohn--Sham independent-particle (IPA@KS) level for high-lying transitions (Supplemental Material, Sec.~S3). The pseudopotential/basis representation of high-lying unoccupied states therefore contributes to the residual intensity differences among codes.

We notice that an overall redshift of the calculated spectrum is observed compared with the experiment. As we have explained in Sec.~\ref{sec:k}, this discrepancy can be attributed to the fact that the $G^0W^0$ underestimates the band gap (7.27~eV vs.\ 7.98~eV in MgO case). The residual redshift is therefore a consequence of the underestimation of the $GW$ quasiparticle energy gap. This can be mitigated with sc-$GW$ or QS$GW$ \cite{jia2026}. These self-consistent calculations are beyond the scope of this paper, while the present benchmark focuses on validating the BSE implementation at the $G^0W^0$ level, which is widely used in practice. The extension of the present BSE implementation to a preceding self-consistent $GW$ reference calculation is relatively straightforward. 

The Si and MgO benchmark results demonstrate the correctness of the BSE implementation within ABACUS+LibRPA for both small-gap covalent semiconductors and wide-gap ionic insulators. This, together with the previous benchmark on Thiel's molecular set, confirms the validity of ABACUS+LibRPA BSE functionality across different materials.

\subsubsection{Exciton binding energies}

Beyond the spectral benchmarks, the exciton binding energies provide a sensitive probe of material-dependent sampling requirements and further motivate the dual-$\bm k$-mesh strategy. It is defined as 
\begin{equation}
	E_B=\min_{\bm k}[E_{c\bm k}^{\mathrm{QP}}-E_{v\bm k}^{\mathrm{QP}}]-\Omega_1^{\mathrm{BSE}},
\end{equation}
the difference between the minimum direct quasiparticle gap and the lowest BSE excitation energy. This definition captures the energy lowering due to the attractive electron-hole interaction relative to the independent-quasiparticle picture. Table~\ref{tab:EB_summary} reports values for six materials, alongside literature results~\cite{Alvertis2023} and experimental data.

\begin{figure}[hbtp]
\centering
\includegraphics[width=0.40\textwidth]{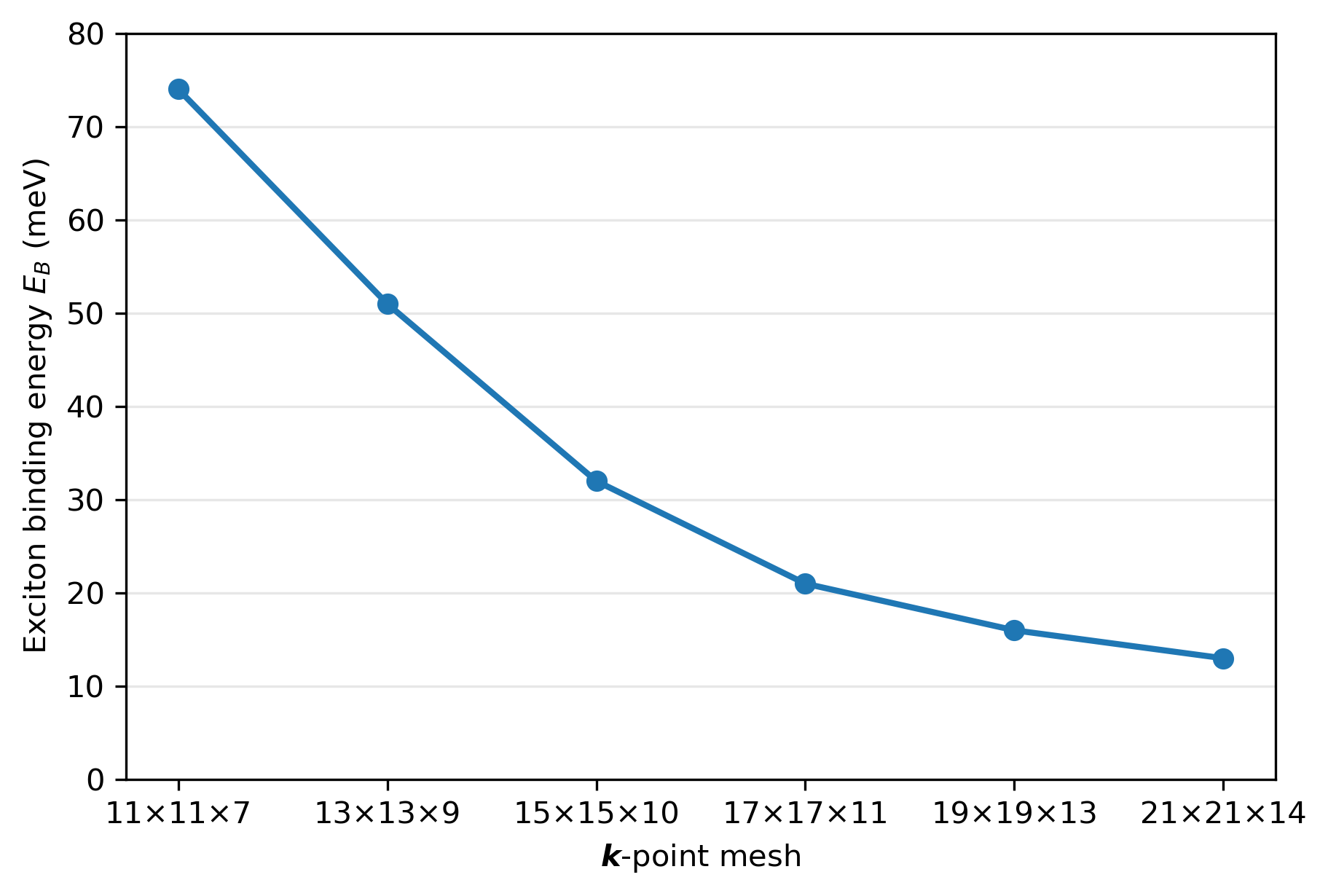}
\caption{GaN exciton binding energy $E_B$ as a function of $\bm k$-point mesh. The values decrease from 74~meV ($11\times11\times7$) to 32~meV ($15\times15\times10$) and 13~meV ($21\times21\times14$), illustrating the strong mesh dependence.}
\label{fig:GaN_EB_kmesh}
\end{figure}
\onecolumngrid
\vfill
\noindent\begin{minipage}{\textwidth}
\begin{table}[H]
\caption{Exciton binding energies $E_B$ (meV) for representative materials.}
\label{tab:EB_summary}
\begin{ruledtabular}
    \begin{tabular}{llllllll}
    Material          & Identifier~\cite{Jain2013MP} & $N_\text{occ}$ & $N_\text{virt}$ & $E_B$ (this work) & $E_B^\text{uniform}$ (Ref.~\cite{Alvertis2023}) & $E_B^\text{nonuniform}$(Ref.~\cite{Alvertis2023}) & $E_{B,\text{exp.}}$ \\
    \hline
    AlN (Wurtzite)    & mp-661  & 5 & 5 & 85 (20×20×12)  & 184 (24×24×12) & 147& 48~\cite{AlN-1}, 80~\cite{AlN-2}   \\
    CdS (Zinc blende) & mp-2469 & 4 & 4 & 29 (21×21×21)  & 65 (28×28×28)  & 39 & 28~\cite{CdS-1}, 30~\cite{CdS-2}   \\
    GaN (Wurtzite)    & mp-804  & 6 & 8 & 13 (21×21×14)  & 111 (24×24×12) & 65 & 20~\cite{GaN-1}, 28~\cite{GaN-2}   \\
    MgO (Halite)      & mp-1265 & 4 & 5 & 308 (21×21×21) & 360 (24×24×24) & 323& 80~\cite{MgO-1}, 145~\cite{MgO-2}  \\
    Si (Diamond)      & mp-149  & 4 & 4 & 17 (21×21×21)  & 44 (20×20×20)  & 25 & 15~\cite{Si}                      \\
    SnO$_2$ (Rutile)  & mp-856  & 4 & 8 & 102 (18×18×27) & 124 (18×18×27) & 107& 33~\cite{SnO2-1}, 35~\cite{SnO2-2} \\
    \end{tabular}
\end{ruledtabular}
\end{table}
\end{minipage}
\newpage
\twocolumngrid

The magnitude of the finite $\bm k$-sampling effect is material dependent. For GaN (shown in Figure~\ref{fig:GaN_EB_kmesh}), $E_B$ decreases from 74 to 13~meV as the mesh is refined from $11\times11\times7$ to $21\times21\times14$, indicating pronounced sensitivity to reciprocal-space resolution. In the Wannier--Mott picture, a spatially extended exciton is localized in reciprocal space and therefore requires dense sampling. By contrast, MgO shows much smaller relative changes over the tested mesh range (341$\rightarrow$308~meV from $11\times11\times11$ to $21\times21\times21$). The relatively large $E_B$ is consistent with a smaller exciton Bohr radius and stronger electron-hole binding, which leads to a more localized exciton in real space and a more delocalized wave function in reciprocal space.

The exciton binding energies presented in Table~\ref{tab:EB_summary} are obtained at the limit of our current computational resources, and for several materials the exciton binding energy is not yet fully converged. Achieving higher convergence with uniform $\bm k$-mesh would require a cubic increase in the number of $\bm k$-points, leading to a substantial increase in computational cost. Alvertis \textit{et al.}~\cite{Alvertis2023} demonstrated that nonuniform BZ sampling strategies, in which a dense $\bm k$-mesh is employed only in a localized patch around the exciton center while a coarser mesh is used elsewhere, can achieve fully converged $E_B$ values at a fraction of the computational cost of uniform sampling. Developing such a nonuniform sampling strategy within our computational framework is the subject of ongoing work.

Compared with the experiment, the narrow-gap compounds, like GaN (13~meV), Si (17~meV) and CdS (29~meV) are close to the experimental results (20--28~meV, 15~meV and 28--30~meV). Larger residual deviations remain for wide-gap compounds, notably MgO (308~meV vs.\ 80--145~meV) and SnO$_2$ (102~meV vs.\ 33--35~meV), where gap accuracy, dynamical screening beyond the static approximation, and temperature-dependent phonon screening effects may have a stronger impact on $E_B$. Rigorous convergence of $E_B$ with respect to $\bm k$-point sampling is a critical prerequisite before addressing these additional physical effects, since the magnitude of the sampling correction can be comparable to or larger than the corrections from dynamical screening and electron-phonon coupling.

\section{Conclusion and outlook} \label{sec:Conclusion}

In this work, we have presented a $GW$+BSE implementation within the ABACUS+LibRPA framework that addresses the $\bm k$-sampling bottleneck through the dual-$\bm k$-mesh strategy. The implementation is formulated in NAO basis sets combined with norm-conserving pseudopotentials, and exploits the LRI technique for a compact treatment of the screened interaction. The atom-centered representation labels the screened interaction by real-space lattice vectors and enables its Fourier interpolation to denser $\bm k$-meshes, while enabling independent refinement of the coarse $\bm k$-mesh $GW$ input and fine $\bm k$-mesh transition space.

Benchmark calculations on Thiel's set of 28 molecules demonstrate close agreement with FHI-aims across all four channels (singlet/triplet, TDA/full BSE), with MADs ranging from 32.0 to 43.9~meV. For periodic systems, the computed absorption spectra of Si and MgO are consistent with those from established codes (BerkeleyGW, VASP, Exciting, and FHI-aims) as well as with experimental data. We further analyze the convergence behavior with respect to the NAO basis, the auxiliary basis, the $\bm k$-mesh density, and offset averaging; the offset-averaging scheme mitigates sampling artifacts and mesh-offset sensitivity, but does not by itself establish $\bm k$-point convergence.

These results establish the dual-$\bm k$-mesh strategy as a viable approach for BSE calculations in both molecular and periodic systems within the NAO-plus-pseudopotential framework.

The benefit of the NAO framework depends on the stage of the calculation. Its main computational advantage lies in the preceding $GW$ step, where the real-space sparsity of NAOs and LRI supports low-scaling response and self-energy calculations~\cite{Zhang2026,gong2026}. This is particularly favorable for large unit cells and for systems containing vacuum, such as surfaces, interfaces, and low-dimensional materials. At the BSE stage, the localized representation enables the coarse-to-fine interpolation of the interaction kernel, whereas the dimension of the BSE matrix is set by the transition space and is not reduced by the choice of basis. The trade-off is that the completeness of both the orbital and auxiliary bases must be checked explicitly, as done here through the velocity--length comparison and the PCA-threshold tests.

Future work will target symmetry-compatible real-valued BSE matrix representations, following approaches used in BerkeleyGW~\cite{Deslippe2012}, and Haydock recursion~\cite{Haydock1980} for absorption spectra to reduce computational and memory costs. We also plan to develop nonuniform $\bm k$-mesh interpolation to accelerate spectral convergence and finite-momentum ($\bm q\neq 0$) extensions for momentum-resolved spectroscopy.

\begin{acknowledgments}
    We thank Rong Shi and Songrui Liu for helpful discussions.
    	We acknowledge the funding support from the National Key Research and Development Program of China (Grant Nos.2023YFA1507004 and 2022YFA1403800) and the Strategic Priority Research Program of the Chinese Academy of Sciences under Grant No. XDB0500201. This work was also supported by the National Natural Science Foundation of China (Grants Nos. 12134012, 12374067, and 12188101).
\end{acknowledgments}

\appendix
\section{BSE Spectrum in the long wavelength limit} \label{sec:spectrum derivation}
In this appendix, we briefly derive the formula for the absorption spectrum of BSE. The light absorption capability of a material can be described by the imaginary part of the macroscopic dielectric function in the long-wavelength limit $\varepsilon_2(\omega) = \operatorname{Im} \varepsilon_M(\bm q\rightarrow 0, \omega)$. The macroscopic dielectric function is defined by the $\bm G=0$ component in the plane-wave representation
\begin{equation}\label{macro}
	\varepsilon_M(\bm q,\omega) = \frac{1}{\varepsilon_{\bm G=0,\bm G'=0}^{-1}(\bm q,\omega)}\, ,
\end{equation}
where $\varepsilon_{\bm G=0,\bm G'=0}^{-1}$  is the head element (${\bm G = G'=0}$) of the inverse of the dielectric matrix given by
\begin{equation}\label{diele}
	\varepsilon_{\bm{GG^{\prime}}}^{-1}(\bm{q},\omega)=\delta_{\bm{GG^{\prime}}}+\frac{4\pi}{|\bm{q}+\bm{G}|^2}\chi_{\bm{GG^{\prime}}}(\bm{q},\omega)\, .
\end{equation}
Here $\chi_{\bm{GG^{\prime}}}(\bm{q},\omega)$ is the full interacting response matrix, which is associated with Eq.\eqref{eq:L_matrix1} as
\begin{equation}
	\chi_{\bm G=0,\bm G'=0}(\bm{q},\omega)=\int \mathrm d \bm r_1 \mathrm d \bm r_2 e^{i\bm q\cdot (\bm r_2-\bm r_1)}L(\bm r_1\bm r_2;\bm r_1\bm r_2;\omega)\, .
\end{equation}

As $\bm q \to 0$ and $\bm G=\bm G'=0$, both the numerator and the denominator of Eq.~\eqref{diele} diverge. The viable remedy is to evaluate the expressions at finite $\bm q$ and then take the $\bm q\to 0$ limit. The plane-wave representation is intimately connected to the Bloch wave-vector index, and thus, we make the Bloch wave vectors explicit: for an electron-hole pair $\psi_{a}^{\bm k +\bm q} \psi_{i}^{\bm{k}*}$, the dipole matrix element reads
\begin{equation}
	\langle \psi_{i}^{\bm k}|e^{-i\bm q\cdot \bm r}|\psi_{a}^{\bm k +\bm q}\rangle = \langle u_{i}^{\bm k}|u_{a}^{\bm k+ \bm q}\rangle \, ,
\end{equation}
where $u_{i}^{\bm k}(\bm r) = \psi_{i}^{\bm k}(\bm r) e^{-i\bm k \cdot \bm r}$ is the periodic part of the Bloch wave function. Using perturbation theory, the wave function at the $\bm k+\bm q$ point can be expanded as
\begin{equation}
	|u_{a}^{\bm k+\bm q}\rangle = |u_{a}^{\bm{k}}\rangle + \sum_{m\neq a} \frac{\langle u_{m}^{\bm k}|\hat V|u_{a}^{\bm k}\rangle }{E_a - E_m} |u_m\rangle \, ,
\end{equation}
where the perturbation Hamiltonian is
\begin{equation}\label{pert}
	\begin{aligned}
    \hat V &=\lim_{\bm q \to 0} \hat H(\bm{k}+\bm{q}) - \hat H(\bm k) \\
    &=\lim_{\bm q \to 0} e^{-i\bm q\cdot \bm r}\hat H(\bm{k})e^{i\bm q\cdot \bm r} - \hat H(\bm k) \\
    &=\lim_{\bm q \to 0} i\bm q \cdot [\hat H, \hat{\bm r}] = \lim_{\bm q \to 0} \bm q \cdot \hat{\bm v} \, .
	\end{aligned}
\end{equation}
Combining Eqs.~\eqref{macro}--\eqref{pert} and using the identity $\operatorname{Im}\!\left[\lim_{\eta\to0}1/(x+i\eta)\right]=-\pi\delta(x)$, one directly obtains the velocity form of Eq.~\eqref{eq:abs_vel}. We note that the plane-wave functions are normalized in the Born--von~K\'arm\'an supercell, which introduces the additional factor $N_{\bm k}V$ in the denominator.

At this point, we have completed the derivation for the $\bm q=0$ case, which is the focus of our current work. In the BSE Hamiltonian matrix, different $\bm q$ components are decoupled and can be solved independently. Finite-momentum response and energy-loss spectra have been studied in earlier first-principles work~\cite{Weissker2006,Lawler2008,Weissker2010}, as well as in recent BSE developments~\cite{Alam2025}. Such an extension lies beyond the scope of the present work and is deferred to future work.

\section{Comment on length form of absorption for extended state}\label{sec:length form}
As discussed in the main text, the length form is generally not suitable for extended states. A common argument is that the position operator is ill-defined for such states and should be treated within a Berry-phase formalism~\cite{Resta1997,Yaschenko1998}. In the BSE context, however, the issue is different, because Kohn-Sham states are orthogonal, the transition dipoles between occupied states and unoccupied states $\langle i\bm{k}|\hat{\bm r}|a\bm{k}\rangle$ are independent of the choice of coordinate origin.

The key point is instead related to operator domains and Hermiticity in extended-state Hilbert spaces: when the position operator acts on a Bloch-like state, the resulting state does not, in general, remain in the domain where the Hamiltonian is Hermitian. Consequently,
\begin{equation}
	\langle \psi_1|\hat H\hat{\bm r}|\psi_2\rangle
	\neq
	\langle \hat H\psi_1|\hat{\bm r}|\psi_2\rangle\, ,
\end{equation}
and therefore the matrix element of the length form is not directly related to the velocity form, which means the following inequality
\begin{equation}
	\langle  \psi_1|\hat{\bm r}|\psi_2\rangle \neq
	i \frac{\langle \psi_1|\hat{\bm v}|\psi_2\rangle}{E_2-E_1}\, .
\end{equation}
A more detailed discussion can be found in Sec. 2.4 of Ref.~\cite{Esteve2023}, where they derive the surface term and explain it as the source of difference.
Here we illustrate this point by a one-dimensional free particle. Under box normalization on the interval $[0,L]$, consider the two plane-wave states
$\psi_1(x)=e^{i\frac{2\pi}{L}x}/\sqrt{L}$ and $\psi_2(x)=e^{i\frac{4\pi}{L}x}/\sqrt{L}$, we have
\begin{equation}
	\begin{aligned}
		\langle\psi_1|\hat H\hat x|\psi_2\rangle
		 & = \frac{1}{L}\int_{0}^{L}\mathrm{d}x\,e^{-i\frac{2\pi}{L}x}
		\left(-\frac{1}{2}\pdv[2]{x}\right) x e^{i\frac{4\pi}{L}x} \\
		 & = -i\frac{4\pi}{L}\, , \\
		\langle\hat H\psi_1|\hat x|\psi_2\rangle
		 & = \frac{2\pi^2}{L^3}\int_{0}^{L}\mathrm{d}x\,x e^{i\frac{2\pi}{L}x}
		= -i\frac{\pi}{L}\, ,
	\end{aligned}
\end{equation}
The mismatch between these two formulae can be seen from the surface term during the process of integration by parts
\begin{equation}
   -\frac 1 2  \left.\left[ \psi_1^* \pdv{(x\psi_2)}{x} - \pdv{\psi_1^*}{x}x\psi_2 \right] \right|_0^L = -i \frac{3\pi}{L}\, .
\end{equation}

For bound states, the surface term vanishes as the integration boundary is pushed to infinity where the wave function decays to zero. For Bloch states, the surface term at opposite boundaries of the Born–von Kármán supercell cancels due to the periodic boundary condition. In both cases, the Hamiltonian remains Hermitian, and the velocity and length gauges are formally equivalent. However, once the position operator is applied to a Bloch function, the resulting state no longer respects the periodicity of the supercell, and the surface term persists. This explicitly demonstrates why the velocity and length forms are not generally equivalent for extended states.
\nocite{*}
\bibliography{aipsamp}

\end{document}

%% file: newcommands.tex
\definecolor{mod}{RGB}{1, 113, 192}

%% file: fig/workflow.tex
\begin{tikzpicture}[
    font=\Large,
    >=Stealth,
    arr/.style={->, line width=0.8pt},
    redbox/.style={
        draw=red,
        line width=1.2pt,
        fill=white,
        inner sep=6pt,
        align=center
    },
    bluebox/.style={
        draw=cyan,
        line width=1.2pt,
        fill=white,
        inner sep=6pt,
        align=center
    },
    redbluebox/.style={
        draw=none,
        line width=2pt,
        fill=none,
        inner sep=6pt,
        outer sep=0pt,
        align=center,
        path picture={
            \draw[red,line width=2pt] (path picture bounding box.north west) -- (path picture bounding box.north east); 
            \draw[cyan,line width=2pt] (path picture bounding box.north west) -- (path picture bounding box.south west); 
            \draw[cyan,line width=2pt] (path picture bounding box.north east) -- (path picture bounding box.south east); 
            \draw[red,line width=2pt] (path picture bounding box.south west) -- (path picture bounding box.south east); 
        }
    },
    bigbox/.style={
        draw=black,
        line width=0.8pt,
        rounded corners=10pt,
        fill=white
    }
]


\node[bigbox, minimum width=13cm, minimum height=3.2cm, anchor=north]
    (abacus1) at (0,0) {};
\node[anchor=north east, align=right, font=\huge] at (6.4,-0.25) {ABACUS};

\node[bigbox, minimum width=13cm, minimum height=8.4cm, anchor=north]
    (librpa) at (0,-3.45) {};
\node[anchor=north east, align=right, font=\huge] at (6.4,-3.80)
    {LibRPA\\(LibRI)};

\node[bigbox, minimum width=13cm, minimum height=9.2cm, anchor=north]
    (abacus2) at (0,-12.15) {};
\node[anchor=north east, align=right, font=\huge] at (6.4,-12.45)
    {ABACUS\\(LibRI)};


\node[draw=black, line width=0.8pt, minimum width=8.0cm, minimum height=1cm,
      font=\huge, align=center]
    (input) at (0,1.1) {INPUT, STRU, KPT};


\node[redbox] (V) at (-2.0,-1.05)
    {$V_{\mu\nu}(\bm{R}),\; V_{\mu\nu}(\bm{q})$};

\node[redbox, below=0.45cm of V] (C)
    {$C_{st}^{\mu}(\bm{R})$};

\node[redbluebox] (E) at (0.85,-1.35)
    {$E_n^{\bm{k}}$};

\node[redbluebox] (c) at (2.25,-1.35)
    {$c_{sn}^{\bm{k}}$};

\draw[arr] (input.south) -- (V.north);
\draw[arr] (input.south) -- (E.north);
\draw[arr] (input.south) -- (c.north);

\draw[arr] (V.south) -- (C.north);


\node[redbox] (G) at (0,-4.65)
    {$G_{st}^{0}(\bm{R}, i\tau)$};

\node[redbox, below=0.5cm of G] (chi)
    {$\chi_{\mu\nu}^{0}(\bm{R}, i\tau),\;
      \chi_{\mu\nu}^{0}(\bm{q}, i\omega)$};

\node[redbox, below=0.5cm of chi] (W)
    {$W_{\mu\nu}(\bm{q}, i\omega),\;
      W_{\mu\nu}(\bm{R}, i\tau)$};

\node[redbox, below=0.5cm of W, xshift=1.5cm] (SigmaR)
    {$\Sigma_{st}(\bm{R}, i\tau)$};
\node[bluebox, below=0.5cm of W, xshift=-1.5cm] (Sigmak)
    {$\Sigma_{st}(\bm{k}, i\omega)$};

\node[bluebox] (EGW) at (-2,-10.70)
    {$E_{n\bm{k}}^{\mathrm{GW}}$};

\node[redbox] (Wzero) at (1,-10.70)
    {$W_{\mu\nu}(\bm{R}, i\omega \approx 0)$};

\draw[arr] (E.south) -- ++(0,-0.5) to[bend right=90] ++(0,-0.3) -- (E.south |- G.north);
\draw[arr] (E.west) -- ++(-0.25,0) -- ++(0,-1) to[bend right=90] ++(0,-0.3) |- ($ (EGW.west)+(-0.4,7) $) 
    -- ++(0,-2.35) to[bend right=90] ++(0,-0.3) -- ++(0,-1.25) to[bend right=90] ++(0,-0.3) 
    -- ++(0,-1.8) -- (EGW.north);
\draw[arr] (c.south) -- ++(0,-0.5) to[bend right=90] ++(0,-0.3) |- (G.east);

\draw[arr] (G.south) -- (chi.north);
\draw[arr] (chi.south) -- (W.north);
\draw[arr] (W.south) -- (SigmaR.north);
\draw[arr] (SigmaR.west) -- (Sigmak.east);
\draw[arr] (Sigmak.south) -- (EGW.north);

\draw[arr] (W.east) -- ++(0.45,0) |- (Wzero.east);

\draw[arr] (C.west) -- ++(-0.5,0) |- (chi.west);
\draw[arr] (V.west) -- ++(-0.5,0) |- (W.west);


\node[bluebox] (Ctilde) at (0,-13.25)
    {$\widetilde{C}^{\mu}_{j\bm{k}_{2};i\bm{k}_{1}}$};

\node[bluebox, below=0.5cm of Ctilde] (VWmat)
    {$
    \begin{aligned}
    &(a\bm{k}_{1}^{*}, i\bm{k}_{1}|V|j\bm{k}_{2}^{*}, b\bm{k}_{2})\\
    &(j\bm{k}_{2}^{*}, i\bm{k}_{1}|W|a\bm{k}_{1}^{*}, b\bm{k}_{2})
    \end{aligned}
    $};

\node[bluebox, below=0.5cm of VWmat] (HBSE)
    {$H^{\mathrm{BSE}}_{ai\bm{k}_{1},bj\bm{k}_{2}}$};

\node[bluebox, below=0.5cm of HBSE] (Omega)
    {$\Omega^{S},\; Z^{S}_{ai\bm{k}_{1}}$};

\node[bluebox, below=0.5cm of Omega] (eps)
    {$\varepsilon_{2}(\omega)$};

\draw[arr] (Ctilde.south) -- (VWmat.north);
\draw[arr] (VWmat.south) -- (HBSE.north);
\draw[arr] (HBSE.south) -- (Omega.north);
\draw[arr] (Omega.south) -- (eps.north);

\coordinate (Wzoff) at ($ (Wzero.west)+(-0.25,0) $);
\draw[arr] (Wzero.west) -- (Wzoff) -- (Wzoff |- VWmat.north);
\draw[arr] (c.east) -- ++ (0.6,0) |- (Ctilde.east);
\coordinate (mid) at ($ (c.east) + (0.6,-1.15) $);
\draw[arr] (C.east) -- (mid) |- (Ctilde.east);
\draw[arr] (EGW.west) -- ++(-0.4,0) -- ++(0,-4.45) to[bend right=90] ++(0,-0.3) |- (HBSE.west);
\draw[arr] (V.west) -- ++(-0.9,0) |- (VWmat.west);

\end{tikzpicture}